\documentclass[sigconf]{acmart}

\AtBeginDocument{%
  }

\usepackage{stfloats}
\usepackage{float}
\usepackage{placeins}
\usepackage{graphicx}
\usepackage{subcaption}
\usepackage{enumitem}
\usepackage{tabularx}
\usepackage{booktabs}
\usepackage{listings}

\definecolor{promptbg}{RGB}{245,245,245}
\definecolor{promptframe}{RGB}{180,180,180}
\definecolor{promptkw}{RGB}{0,80,160}
\definecolor{promptstr}{RGB}{160,60,0}

\lstdefinestyle{prompt}{
  basicstyle=\ttfamily\footnotesize,
  breaklines=true,
  breakindent=0pt,
  columns=fullflexible,
  frame=single,
  rulecolor=\color{promptframe},
  backgroundcolor=\color{promptbg},
  showstringspaces=false,
  upquote=true,
  aboveskip=6pt,
  belowskip=6pt,
  xleftmargin=4pt,
  xrightmargin=4pt,
  framexleftmargin=4pt,
  framexrightmargin=4pt,
  tabsize=2,
  keepspaces=true,
}

\author{Lana Do}
\authornote{Corresponding author.}
\affiliation{%
  \institution{Northeastern University}
  \country{United States}
}
\email{do.ng@northeastern.edu}

\author{Shasta Ihorn}
\affiliation{%
  \institution{San Francisco State University}
  \country{United States}
}
\email{sihorn@sfsu.edu}

\author{Charity M. Pitcher-Cooper}
\affiliation{%
  \institution{The Smith-Kettlewell Eye Research Institute}
  \country{United States}
}
\email{cpc@ski.org}

\author{Sanjay Mirani}
\affiliation{%
  \institution{San Francisco State University}
  \country{United States}
}
\email{miranisanjay417@gmail.com}

\author{Gio Jung}
\affiliation{%
  \institution{San Francisco State University}
  \country{United States}
}
\email{gjung1@sfsu.edu}

\author{Hyunjoo Shim}
\affiliation{%
\institution{San Francisco State University}
\country{United States}
}
\email{shim.hy@northeastern.edu}

\author{Zhenzhen Qin}
\affiliation{%
  \institution{Northeastern University}
  \country{United States}
}
\email{qin.zhen@northeastern.edu}

\author{Kien T. Nguyen}
\affiliation{%
  \institution{Northeastern University}
  \country{United States}
}
\email{nguyen.trungk@northeastern.edu}

\author{Vassilis Athitsos}
\affiliation{%
  \institution{University of Texas at Arlington}
  \country{United States}
}
\email{athitsos@uta.edu}

\author{Ilmi Yoon}
\affiliation{%
  \institution{Northeastern University}
  \country{United States}
}
\email{yoon.i@northeastern.edu}

\begin{document}


\begin{abstract}
Audio description (AD) narrates visual information in video for blind and low-vision audiences. AI-generated drafts can lower the barrier to AD authoring for novice describers, but whether drafts of differing quality provide the same benefits remains an open question. In a within-subjects study, 30 novice describers authored AD from scratch and edited unguided and accessibility-guideline-informed AI drafts using an instrumented interface. We also introduce the Multi-Dimensional Contribution Index (MDCI), which measures human and AI contribution across text, timing, playback, and delivery. Higher-quality drafts reduced completion time and workload, while lower-quality drafts prompted more deletion and major rewriting. Across conditions, editors also restructured clips, added missing information, adjusted timing, and adapted descriptions to the content, reflecting contributions beyond textual revision. However, novice describers did not reliably resolve all problems in AI drafts: existing wording could anchor subsequent edits, and subtle inaccuracies were not always corrected. Together, these findings suggest that effective AI-assisted authoring depends not only on providing a useful starting point, but also on understanding how that starting point shapes subsequent human effort, judgment, and contribution.
\end{abstract}



\title{How Do Novice Describers Edit AI Drafts? The Role of Draft Quality in Audio Description Authoring}

\acmConference[IUI '27]{32nd International Conference on Intelligent User Interfaces}{}{}

\maketitle

\section{Introduction}

Audio description (AD) conveys essential visual information through spoken narration, making video content accessible to blind and low-vision (BLV) audiences. Professional AD is costly and time-consuming
\cite{branje2012livedescribe}, leaving the growing volume of user-generated and short-form video content largely undescribed
\cite{Pitcher-Cooper2023YouDataset}. Multimodal Large Language Models (MLLMs) have opened a path to generating AD at scale
\cite{Wang2021TowardVideos, chu2024llmad, Ye2024MMAD:Description}, but
AI-produced descriptions still suffer from hallucinations, verbosity, and lack of contextual coherence \cite{li2025vidhalluc, ihorn2023potential}. Human oversight remains necessary. On volunteer platforms, audio description quality stewardship falls largely to novice describers rather than trained professionals \cite{snyder2014visual}, with volunteers serving as the primary contributors bringing accessible content to BLV audiences. Designing effective editing environments for volunteer authors is not straightforward: AD editing requires visual judgment, temporal reasoning, and domain knowledge of accessibility conventions
\cite{Pavel2020Rescribe:Descriptions, liu2022crossa11y, natalie2023supporting}.

Prior work demonstrates that human-AI workflows for AD are viable and can lower the barrier to entry for novice describers. Yuksel et al.\ showed that starting from an AI-generated draft was faster and easier for first-time volunteers on YouDescribe than working without AI assistance, even when describer-rated draft accuracy was only neutral
\cite{Yuksel2020Human-in-the-LoopUsers}. Cheema et al.\ extended this with collaborative co-authoring that reduces repetitive effort and eases cognitive load \cite{cheema2025describepro}. In practice, however, MLLM output is highly sensitive to prompting strategy: minimal prompts yield verbose, caption-like text, while guideline-informed prompts produce descriptions BLV users rate comparably to those of trained describers \cite{chen2024sharegpt4video, lin2023mmvid, Li2025VideoA11y:Description, xie2024autoad}.

This raises a question prior work has not addressed: how does draft quality shape the editing process? Specifically, how does it affect where describers direct their editing effort, what content they retain or remove, and how they experience the task? Prior studies have either held draft quality fixed~\cite{Yuksel2020Human-in-the-LoopUsers, cheema2025describepro} or evaluated quality only through ratings of the final output~\cite{Li2025VideoA11y:Description}.
Related work in adjacent domains shows that draft quality directly shapes editing effort. In machine translation post-editing, low-quality output can cost editors more than translating from scratch and can anchor them to flawed text ~\cite{Green2013TheTranslation, LaubliAssessingEnvironment}. In AI-assisted writing, the level and configuration of assistance shape what writers produce, how they think, and how they experience the task ~\cite{arnold2020predictive, bhat2023nextphrase, dhillon2024shaping, hemmer2023delegation}. Whether and how these dynamics carry over to AD --- a task with distinct demands of visual interpretation, tight temporal windows, and accessibility conventions --- is an open empirical question.

To study how starting-draft quality shapes human refinement, we use two AI generation approaches that produce meaningfully different starting drafts while holding the editing environment constant. Baseline uses a simple prompt without accessibility-specific guidance, while \textbf{GenAD} incorporates professional AD guidelines~\cite{2024DescribedDCMP, NationalCenterforAccessibleMedia2017AccessibleGuidelines, YouDescribeYouDescribe.Https://www.youdescribe.org/} and contextual video information. Participants edit both types of drafts in \textbf{RefineAD}, an instrumented authoring interface that captures fine-grained editing activity. 
RefineAD also supports a collaborative mode in which multiple contributors can build incrementally on each other's work. Both components were designed in consultation with accessibility experts, including an experienced audio describer who trains novice volunteers and a professional blind quality control specialist. To measure human contribution in this hybrid authoring process, we introduce a \textbf{Multi-Dimensional Contribution Index} (MDCI) that decomposes each editor’s modifications across text, timing, playback, and voice.

In summary, this paper contributes:
\begin{enumerate}
    \item An empirical characterization of how novice describers edit AI-generated AD drafts of differing quality, from a within-subjects study ($N$=30) combining task performance, logged editing behavior, contribution measurement, and qualitative experience.
    \item The Multi-Dimensional Contribution Index (MDCI), a method for quantifying human and AI contribution in hybrid AD authoring across text, timing, playback, and delivery.
    \item GenAD and RefineAD, an accessibility-informed generation pipeline and instrumented editing interface used to study human refinement across AI drafts of differing quality.
\end{enumerate}

\section{Related Work}

\subsection{AI-Driven and Participatory AD Systems}
The scarcity of professional AD and the growth of online video content have motivated automated generation. Wang et al.\ built Tiresias, combining audiovisual inconsistency detection, dense video captioning, and GPT-based optimization to generate and place AD \cite{Wang2021TowardVideos}. LLM-AD used GPT-4V with scene segmentation and character tracking \cite{chu2024llmad}, while MMAD combined CLIP with LLaMA2 to integrate subtitles and sound cues for narrative consistency \cite{Ye2024MMAD:Description}. ShortScribe used BLIP-2 and GPT-4 to generate hierarchical video summaries at multiple levels of detail \cite{VanDaele2024MakingSummaries}. VideoA11y prompted GPT-4V with 42 professional AD guidelines to generate descriptions \cite{Li2025VideoA11y:Description}. However, automated AD alone often falls short. Systems misidentify characters or insert hallucinated details \cite{Wang2021TowardVideos, li2025vidhalluc}, produce verbose or redundant descriptions \cite{ihorn2023potential}, and score well on captioning metrics yet misalign with human judgments of AD quality \cite{Rohrbach2017MovieDescription}. AI-generated descriptions have also been shown to encode societal biases, defaulting to hegemonic assumptions about race, gender, and appearance \cite{bergin2025automating}. Moreover, AD has been characterized as a creative and interpretive practice involving subjectivity and stylistic nuance, qualities that remain difficult to automate \cite{walczak2017creative, schaefferlacroix2023beyond}.

Crowdsourcing offers another path to expanding AD coverage: platforms like YouDescribe let volunteers record descriptions synchronized to YouTube videos \cite{YouDescribeYouDescribe.Https://www.youdescribe.org/}. But volunteer platforms lack the multi-stage workflow of professional
production: drafting by trained describers, iterative editorial and timing review, and studio recording by voice actors \cite{fryer2016introduction, snyder2014visual}. A range of tools have been developed to support participatory AD authoring. In human-first workflows, describers author and AI refines or critiques. Rescribe \cite{Pavel2020Rescribe:Descriptions} and CrossA11y \cite{liu2022crossa11y} reduced authoring barriers through automatic timing edits and cross-modal accessibility detection, respectively, and Natalie et al.\ provided automated feedback on novices' drafts \cite{natalie2023supporting}. Other work extended authoring to 360° video \cite{Chang2022OmniScribe:Videos}, made the process accessible to BLV creators \cite{huh2023avscript}, gave BLV end-users control over generation \cite{Cheema2024DescribeIndividuals}, or let describers steer generation through natural language prompting \cite{cheema2025describepro}. In AI-first, human-refine workflows, the system generates a complete draft that a human edits; Yuksel et al.\ showed on YouDescribe that novices working from AI drafts were significantly faster than those writing from scratch \cite{Bodi2021AutomatedUsers, Yuksel2020Human-in-the-LoopUsers}.

We focus on the AI-first, human-refine configuration for three reasons. First, it already operates at scale on volunteer platforms such as YouDescribe, making its dynamics consequential for current practice \cite{do2025youdescribe}. Second, it targets the bottleneck that most deters novices: the blank page \cite{rose1980rigid, Yuksel2020Human-in-the-LoopUsers}. Third, in this configuration, the AI draft is the starting material for all subsequent human work, so draft quality propagates through the entire authoring process. 

\subsection{From Output Quality to Editing Experience}
\label{sec:related-draft-quality}

AI-first workflows have been evaluated on volunteer platforms before: Yuksel et al.\ deployed machine-generated drafts to first-time YouDescribe describers \cite{Yuksel2020Human-in-the-LoopUsers}, and Cheema et al.\ introduced collaborative co-authoring to spread the effort across contributors \cite{cheema2025describepro}. AI scaffolding has likewise been shown to support novices in other judgment-intensive authoring tasks such as academic peer reviewing~\cite{sun2024reviewflow, neshaei2024peer}.

These workflows depend on AI-generated drafts, but draft quality varies substantially with how the underlying model is prompted. Generic prompts tend to produce verbose, context-poor output resembling captions more than coherent AD \cite{lin2023mmvid, chen2024sharegpt4video}, while incorporating professional AD guidelines into prompts produces descriptions comparable to trained describers \cite{Li2025VideoA11y:Description}, and prompt design alone can reach competitive performance without fine-tuning \cite{xie2024autoad}. However, these studies evaluate the impact of prompting strategies via BLV end-user ratings of output quality; whether these gains improve the editing experience of describers remains unknown. This question is critical for volunteer-driven platforms like YouDescribe, which rely on novice contributors.   

Prior work in adjacent domains shows that draft quality directly shapes editing effort. In machine translation post-editing, higher-quality outputs reduce workload, while low-quality outputs can be counterproductive, sometimes requiring more effort than translating from scratch \cite{Green2013TheTranslation, LaubliAssessingEnvironment}. Different error types impose different correction costs, making aggregate quality scores an incomplete predictor of effort \cite{daems2017identifying}. Editors also risk anchoring on machine output, accepting or lightly modifying flawed text rather than reconstructing it, which shapes both the process and the product of editing \cite{Green2013TheTranslation}. In AI-assisted writing, the level of scaffolding matters: low-level scaffolding offers little benefit over unassisted writing, whereas high-level scaffolding improves quality but can reduce perceived ownership and satisfaction \cite{dhillon2024shaping}. Even lightweight predictive-text suggestions measurably shape what writers produce \cite{arnold2020predictive}, continuously evaluating suggestions imposes its own cognitive cost \cite{bhat2023nextphrase}, and experts edit a draft differently depending on whether they believe it is AI-generated \cite{hazra2026accepted}.

These findings suggest that draft quality may shape not only the amount of editing effort but also the nature of the edits. Which AI-generated content editors preserve, revise, or discard is itself a key measure of effective human--AI collaboration~\cite{schemmer2023reliance}. We investigate how novice describers edit AD drafts of varying quality, focusing not only on how much they edit but also on the types of edits they make. The unique constraints of AD authoring—including interpreting visual scenes, fitting descriptions into limited time windows, and following accessibility conventions—make understanding these editing behaviors particularly important.

\subsection{Measuring Contribution in Human-AI Co-Creation}

As human-AI collaboration extends across the creative process, from composition through revision and refinement, how to attribute contributions between humans and AI remains an open question. Attribution shapes perceptions of credit, ownership, and motivation \cite{kadoma2024ownership}. In AI-assisted programming, developers’ sense of ownership and willingness to disclose AI involvement vary with the system’s level of autonomy \cite{seo2026whose}. In writing, AI authorship disclosure influences how readers perceive the resulting work \cite{nakano2026disclosure}. Moreover, perceived authorship may diverge from actual contribution: writers using AI assistance may claim authorship over text they do not feel ownership of \cite{draxler2024ghostwriter}.

Machine translation research has developed a rich set of methods for measuring post-editing effort. These include edit-distance metrics such as HTER \cite{snover2006ter}, temporal, technical, and cognitive measures of effort \cite{krings2001repairing}, and process-based analyses of how editing unfolds \cite{daems2017identifying, LaubliAssessingEnvironment, koponen2012comparing}. However, these studies consistently show that edit distance is an imperfect proxy for effort: small textual changes can require substantial judgment, while large edits may be largely mechanical \cite{krings2001repairing, koponen2012comparing}. Similar approaches have been adopted in AI-assisted writing, where researchers examine which portions of a document are authored by humans versus AI \cite{clark2018creative, yuan2022wordcraft, lee2022coauthor, lee2024designspace}. Together, these approaches quantify editing activity but provide less insight into what kinds of changes editors make and why.

AD authoring challenges text-centric measures because contributions extend beyond the written description. Describers may reword text, adjust timing relative to dialogue, change playback format, or replace synthetic speech with their own narration. Although invisible to lexical distance metrics, these changes can substantially shape BLV audiences’ experience of AD \cite{Walczak2018VocalPresence, Fernandez-Torne2015TheCatalan}. Attribution becomes further complicated when multiple contributors iteratively refine descriptions \cite{cheema2025describepro, natalie2021viscene}. The Multi-Dimensional Contribution Index (MDCI) introduced in this paper addresses this gap by measuring contribution across the dimensions exposed by AD authoring tools—description text, temporal placement, playback type, and voice delivery—rather than text alone.

\section{System}

GenAD and RefineAD were designed for community-driven AD platforms where amateur volunteers describe videos without professional tools. Retention on such platforms is limited by the cognitive demands of authoring, motivating AI-generated drafts.

\begin{figure*}[t]
  \centering
  \includegraphics[width=0.98\textwidth]{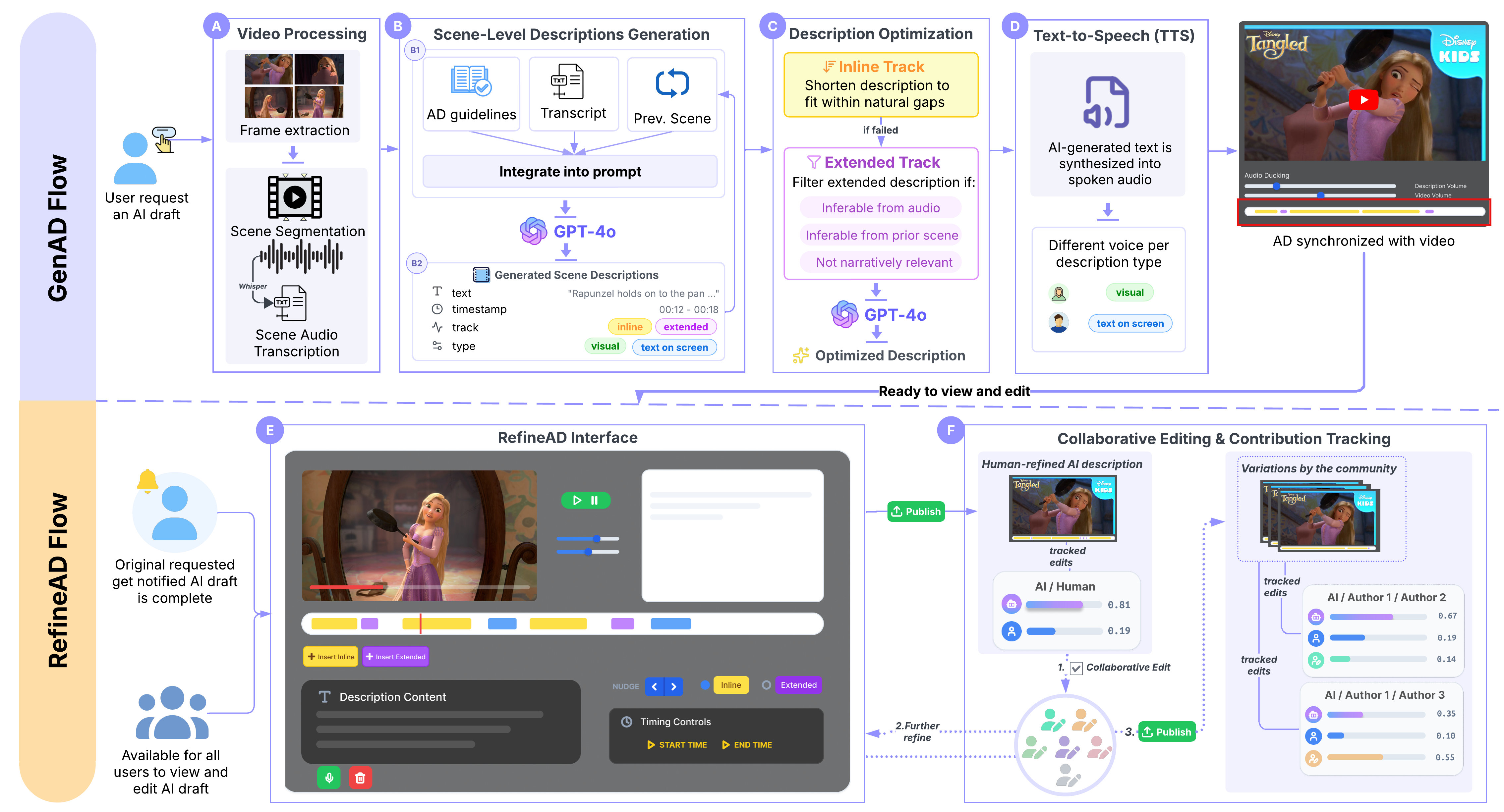}
  \caption{Overview of the GenAD–RefineAD pipeline. GenAD processes a YouTube video through four stages: (A) video processing, (B) scene-level description generation incorporating AD guidelines, audio transcript, and prior scene context, (C) description optimization that prioritizes inline delivery and filters extended descriptions, and (D) text-to-speech synthesis with distinct voices per event type. The output feeds into (E) the RefineAD editing interface, where editors can preview, edit, and publish descriptions, and (F) collaborative editing with multi-author contribution tracking.}
  \label{fig:system}
\end{figure*}

\subsection{GenAD: AD Generation Pipeline}

GenAD generates audio descriptions for YouTube videos on demand, producing drafts that volunteers can preview and refine in RefineAD. Consistent with DCMP guidelines recommending that describers work with familiar content \cite{2024DescribedDCMP}, the system allows users to request any YouTube video, enabling them to begin with material they are familiar with. The user's request triggers a four-stage pipeline: video processing, scene-level generation, description optimization, and text-to-speech (TTS) conversion (Figure~\ref{fig:system}, A–D). Each stage is informed by iterative feedback from a professional audio describer who trains novice volunteers. The output is a complete set of descriptions synchronized with the requested video.

\subsubsection{Video Processing (A)}

As a first step, the pipeline builds two parallel representations of the requested video. On the visual side, frames sampled at roughly 2 fps~\cite{TomarSuramya2006ConvertingFFmpeg} are embedded with CLIP ViT-B/32~\cite{Radford2021CLIP, Schuhmann2022LAION-5B:Models}; consecutive embeddings falling below a cosine similarity of 0.85 (a threshold set empirically during formative testing) mark a scene boundary, yielding visually coherent segments. On the audio side, \texttt{whisper-timestamped} aligns Whisper \texttt{large-v3} transcriptions to word-level timings via dynamic time warping over cross-attention weights~\cite{lintoai2023whispertimestamped,Radford2022RobustSupervision, JSSv031i07}.

\subsubsection{Scene-Level Generation (B)}
Early prototyping revealed that prompting the MLLM (GPT-4o~\cite{openai2024gpt4o}) with minimal instruction (e.g., \textit{``describe the scene''}) produced caption-like outputs rather than coherent descriptions. These were verbose, lacked narrative continuity, and repeated information inferable from dialogue. Our consultants recognized these as common mistakes by novice describers who \textit{``try to do too much.''} To address this, GenAD integrates three inputs (Figure~\ref{fig:system}, B1): professional guidelines that emphasize accurate and descriptive yet concise narration~\cite{2024DescribedDCMP, NationalCenterforAccessibleMedia2017AccessibleGuidelines, YouDescribeYouDescribe.Https://www.youdescribe.org/}, the scene's audio transcript, and accumulated descriptions from prior scenes (full prompt in Appendix~A). Each scene produces structured output (Figure~\ref{fig:system}, B2) with precise timestamps, two event types, \textit{Text-on-Screen} and \textit{Visual}, and track assignment (\textit{Inline} or \textit{Extended}). Inline narration fits within natural pauses in dialogue, and extended narration briefly pauses playback to deliver additional detail~\cite{YouDescribeYouDescribe.Https://www.youdescribe.org/}.

\subsubsection{Description Optimization (C)}
Based on our consultants' feedback to minimize playback interruptions, GenAD prioritizes inline over extended descriptions. An optimization pass condenses current extended clips to fit within dialogue gaps, using retry prompts that progressively shorten text while preserving critical details. If inline delivery remains unfeasible, the system applies a filtering prompt to retain only extended clips that describe necessary information — content not inferable from the audio or prior scenes (Figure~\ref{fig:system}, C).

\subsubsection{Text-to-Speech (D)}
Optimized descriptions are synthesized into spoken audio using Google TTS \cite{googleTTS}, with distinct voices for Visual and Text-on-Screen events (Figure~\ref{fig:system}, D). This was informed by our consultants' feedback that BLV users find the repeated phrase \textit{``text on screen''} fatiguing; a distinct voice lets listeners identify the event type without explicit labeling as well as supporting the broader goal of keeping descriptions concise. 

Finally, the audio clips are synchronized with the video for preview and refinement, with inline and extended narration color-coded on the timeline (yellow and purple, respectively), and audio ducking controls balance narration volume with the original audio. 

\subsubsection{Effect of Prompting Strategy on Output Quality}
\begin{figure*}[t]
    \centering
    \begin{subfigure}[t]{0.52\textwidth}
        \centering
        \includegraphics[width=\linewidth]{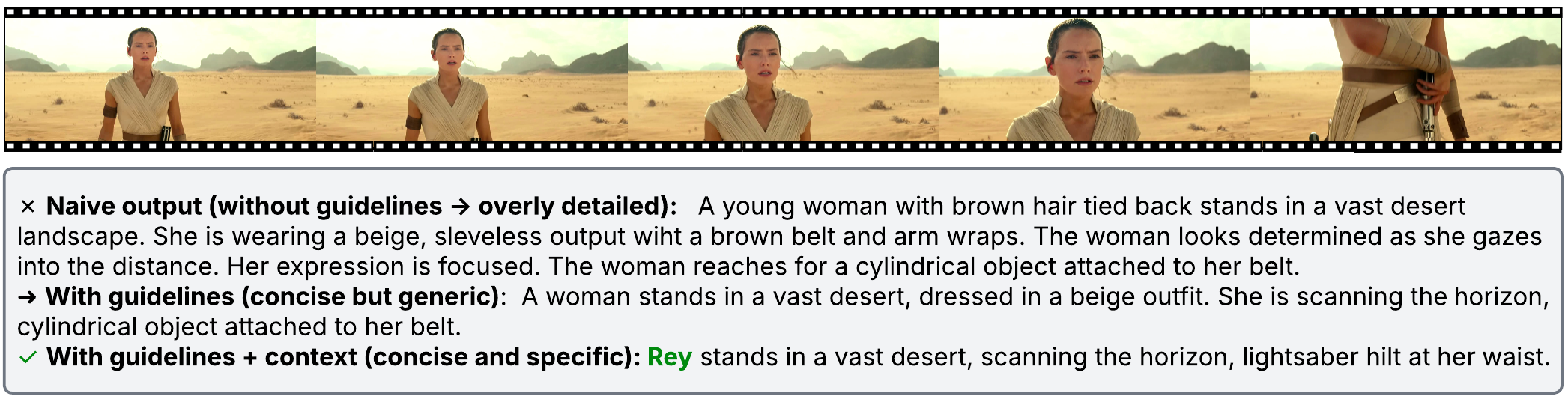}
        \caption{Star Wars Trailer -- progression from simple prompting to guideline-informed and context-aware prompting.}
        \label{starwars-output}
    \end{subfigure}
    \hfill
    \begin{subfigure}[t]{0.45\textwidth}
        \centering
        \includegraphics[width=\linewidth]{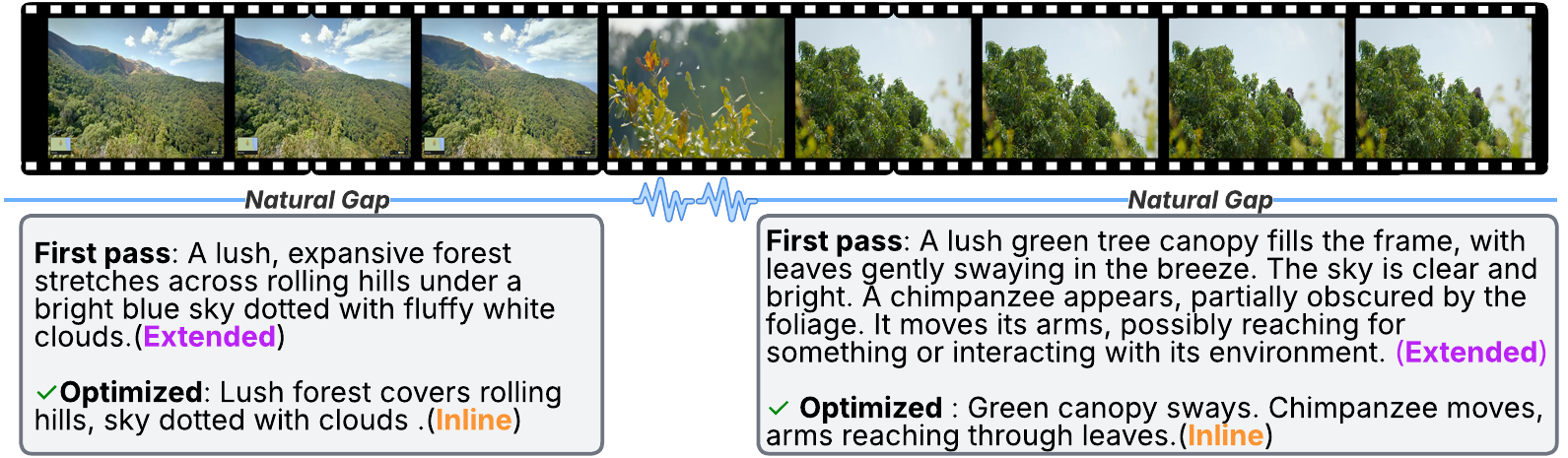}
        \caption{Jane Goodall clip -- optimization condenses extended description into descriptions that fit natural pauses.}
        \label{jane-output}
    \end{subfigure}
    \caption{Qualitative differences between Baseline and GenAD output, shown on videos outside the study corpus.}
    \label{prompting-optimization}
\end{figure*}

We briefly illustrate qualitative differences between the two generation approaches in Figure~\ref{prompting-optimization}. The unguided condition uses only a simple prompt (\textit{``describe the scene''}) without guidelines or context, producing generic outputs like \textit{``a woman stands in the desert.''} With GenAD's full pipeline, the same scene identifies the character by name (Figure~\ref{starwars-output}), and the optimization pass condenses overlong drafts into inline descriptions that fit natural dialogue gaps (Figure~\ref{jane-output}). A quantitative comparison across all study videos is presented in Section~4.

\subsection{RefineAD: Human-AI AD Authoring}
\label{refinead}

\begin{figure*}[t]
  \centering
  \includegraphics[width=\textwidth]
  {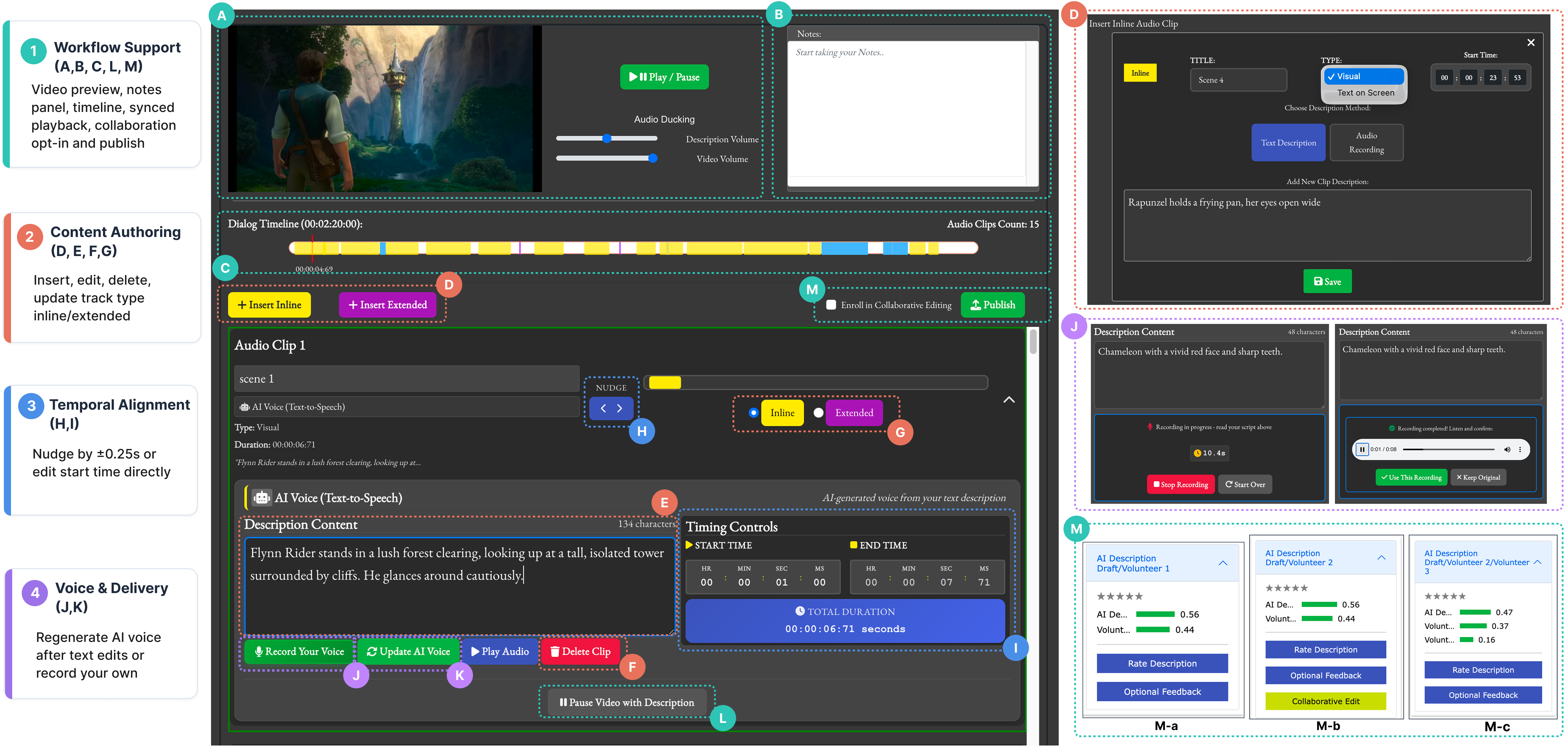}
  \caption{The RefineAD editing interface. Describers preview the video with audio ducking controls (A), take notes (B), and navigate descriptions via a color-coded dialog timeline (C; yellow = inline, purple = extended, blue = dialogue). Editors can insert new inline or extended descriptions (D), edit text (E), delete clips (F), toggle track type (G), adjust timing via $\pm$0.25s nudge or direct entry (H, I), and record a personal voiceover (J) or regenerate the AI voice for updated text (K). Playback synchronizes with the active clip for in-context review (L). Describers publish and can opt into collaborative editing (M), with contribution tracking shown in three modes: (M-a) single-editor view where the original editor has not enabled collaborative editing, (M-b) the editor enables collaborative editing, making the draft available for others to refine, and (M-c) a subsequent collaborator edits the published draft, with contributions tracked across AI, the original editor, and the new collaborator.}
  \label{fig:refinead}
\end{figure*}

Once GenAD completes processing, the original requester is notified via email and linked directly to the editing interface (Figure~\ref{fig:system}, E). The AI draft is also visible to all platform users. If another editor wants to describe the same video, they can begin editing from the preview interface (Appendix~B) without submitting a duplicate request. Each editor's work is saved to their account and can be revisited and edited at any time. All user-facing components, including viewing and editing, were developed and tested with multiple screen readers (JAWS, NVDA, and VoiceOver) and confirmed fully navigable by our blind consultant.
RefineAD preserves the layout of the original freestyle editor (Appendix~C), with preloaded descriptions as the only visible change. GenAD's output serves as editable scaffolding, time-stamped, typed descriptions that editors can accept, modify, or replace. The interface is organized around four aspects of AD editing (Figure~\ref{fig:refinead}).

\subsubsection{Workflow Support}
AD authoring requires describers to simultaneously attend to visual content, dialogue timing, and description quality. To manage this complexity, the interface provides a notes panel for drafting observations during review and a dialog timeline that visualizes all clips alongside dialogue regions. As the video plays with descriptions synchronized, the active clip is highlighted, letting users evaluate descriptions in context rather than in isolation.

\subsubsection{Content Authoring} Users can reshape both the content and structure of the AI draft: inserting new descriptions, editing text, deleting unnecessary clips, or toggling between inline and extended track types. Each clip is pre-labeled by event type (Visual or Text-on-Screen). This design positions AI output as a starting point rather than a finished product, giving users the ability to reshape descriptions based on their own interpretation.

\subsubsection{Temporal Alignment}
Descriptions need to avoid overlapping dialogue and align with the visual events they narrate. Fine-grained $\pm$0.25s nudge controls support incremental adjustment, while direct start and end time editing enables precise placement. The dialog timeline provides visual feedback on whether descriptions fall within or outside dialogue gaps, reducing the trial-and-error typically involved in timing AD.

\subsubsection{Voice and Delivery}
Each clip includes an AI text-to-speech track. After editing text, editors can regenerate the AI voice to reflect their changes or record their own voiceover to replace it entirely. Research shows that BLV audiences often prefer human narration for its clarity and expressiveness, particularly in dramatic or fast-paced content~\cite{Fernandez-Torne2015TheCatalan, Walczak2018VocalPresence}. This supports a range of workflows, from quick edits using AI voice to fully human-narrated descriptions.

\subsection{Contribution Tracking}

When ready to publish, describers can opt into collaborative editing, allowing others to build on their draft. The system tracks contributions across AI and human collaborators, and when collaborative editing is enabled, extends this to multi-author attribution, distinguishing the AI, the original editor, and each subsequent collaborator (Figure~\ref{fig:system}, F).

For each clip that corresponds to an AI-generated clip, we compute a retention score $R \in [0, 1]$:

\begin{equation}
R = w_{\text{text}} \cdot S_{\text{text}} + w_{\text{time}} \cdot S_{\text{time}} + w_{\text{pb}} \cdot S_{\text{pb}} + w_{\text{voice}} \cdot S_{\text{voice}}
\label{eq:retention}
\end{equation}

\noindent $S_{\text{text}}$ measures content preservation across three levels: lexical (normalized Levenshtein \cite{levenshtein1966binary}), semantic (cosine similarity of BGE-M3 embeddings~\cite{chen2024bge}, which encode sentence meaning into dense vectors), and stylistic (cosine similarity of LUAR-MUD embeddings~\cite{rivera2021luar}, which encode authorship patterns such as sentence structure and vocabulary choice). 

\begin{equation}
S_{\text{text}} = w_{\text{lex}} \cdot S_{\text{lex}} + w_{\text{sem}} \cdot S_{\text{sem}} + w_{\text{sty}} \cdot S_{\text{sty}}
\label{eq:text-sim}
\end{equation}

\noindent $S_{\text{time}}$ uses Dynamic Time 
Warping~\cite{sakoe1978dtw}, a sequence alignment method that  measures how much a clip's temporal placement was shifted in the video. $S_{\text{pb}}$ and $S_{\text{voice}}$ are binary indicators for playback type and delivery mode changes. Deleted clips receive $R = 0$; and inserted clips are attributed entirely to the human editor. The overall contribution score is:

\begin{equation}
C_{\text{AI}} = \frac{\sum_{i \in M} w_i \cdot R_i}{\sum_{i \in M} w_i + \sum_{j \in D} w_j + \sum_{k \in I} w_k}, \quad C_{\text{H}_1} = 1 - C_{\text{AI}}
\label{eq:contribution}
\end{equation}

\noindent where $M$, $D$, $I$ are the sets of modified, deleted, and inserted clips; $w_i$ is clip $i$'s word count.

When a subsequent collaborator $\text{H}_2$ edits the $\text{H}_1$-AI version, prior shares scale proportionally:

\begin{equation}
C_{\text{H}_2} = 1 - R', \quad C_{\text{H}_1}' = C_{\text{H}_1} \cdot R', \quad C_{\text{AI}}' = C_{\text{AI}} \cdot R'
\label{eq:multi-author}
\end{equation}

\noindent This extends to any number of collaborators. We weight text and time equally ($w_{\text{text}} = w_{\text{time}} = 0.45$), with playback and voice at $0.05$ each. Within the text dimension: $w_{\text{lex}} = 0.50$, $w_{\text{sem}} = 0.40$, $w_{\text{sty}} = 0.10$. These weights can be further calibrated as more editing data is collected on the platform.

\section{Study Design}
\label{sec:study}

We conducted a within-subjects study with 30 novice describers to evaluate the GenAD-RefineAD pipeline and to test whether AI draft quality affects the editing experience.

\subsection{Study Materials}
\subsubsection{Videos}
We selected five YouTube videos, each approximately 2 minutes long, spanning three genres—instructional, entertainment, and educational—based on requested genres by the BLV community on YouDescribe~\cite{YouDescribeYouDescribe.Https://www.youdescribe.org/}. Two instructional cooking videos, crispy fritters recipe\footnote{\url{https://youtube.com/watch?v=cNj3aOTYdQQ}} and potato balls recipe\footnote{\url{https://youtube.com/watch?v=nqXz8hhAYGo}}, were previously used in Yuksel et al.~\cite{Yuksel2020Human-in-the-LoopUsers}, providing a behavioral anchor to our from-scratch condition on the same content. To increase genre diversity and vary difficulty, we added three new videos: a clip from the animation \textit{Tangled}\footnote{\url{https://youtube.com/watch?v=JXnAd8bofTo}}, a neuroscience lesson\footnote{\url{https://youtube.com/watch?v=6qS83wD29PY}}, and an origami tutorial\footnote{\url{https://youtube.com/watch?v=djPgd1m6IMY}}. 

Based on our consultants' assessment, videos were categorized by difficulty. Cooking instructional and animated videos were classified as \textbf{low difficulty} due to their concrete visual content and sequential actions. Neuroscience and origami videos were classified as \textbf{high difficulty} because they require specialized technical terminology and involve abstract or spatially complex visual content.

\subsubsection{Conditions}
Each participant experienced three conditions across the five videos:
\begin{itemize}
    \item \textbf{From scratch}: No AI draft; participants authored descriptions from a blank editor.
    \item \textbf{Baseline}: An unguided AI draft generated using a minimal prompt (\textit{``describe the scene''}) without guidelines, context, or optimization.
    \item \textbf{GenAD}: The full pipeline output described in Section~3.1.
\end{itemize}

Table~\ref{tab:condition-comparison} summarizes the Baseline and GenAD drafts across all five videos. 
GenAD drafts consistently contained fewer clips overall and relied predominantly on inline playback, reflecting the optimization pass’s consolidation of descriptions into available speech gaps. Baseline drafts, generated without guideline and contextual awareness, produced more clips, a higher proportion of extended descriptions, and a generally higher total word count. 

\begin{table}[!htb]
\centering
\caption{AI-generated draft characteristics across Baseline (B) and GenAD (G) conditions for all five study videos.}
\label{tab:condition-comparison}
\footnotesize
\setlength{\tabcolsep}{3pt}
\begin{tabular*}{\columnwidth}{@{\extracolsep{\fill}}l rr rr rr rr@{}}
\toprule
& \multicolumn{2}{c}{\textbf{Clips}} & \multicolumn{2}{c}{\textbf{Inline}} & \multicolumn{2}{c}{\textbf{Ext.}} & \multicolumn{2}{c}{\textbf{Words}} \\
\cmidrule(lr){2-3} \cmidrule(lr){4-5} \cmidrule(lr){6-7} \cmidrule(lr){8-9}
\textbf{Video} & \textbf{B} & \textbf{G} & \textbf{B} & \textbf{G} & \textbf{B} & \textbf{G} & \textbf{B} & \textbf{G} \\
\midrule
Crispy Fritters & 24 & 24 & 17 & 23 & 7  & 1  & 307 & 298 \\
Potato Balls    & 59 & 35 & 23 & 32 & 36 & 3  & 606 & 315 \\
Tangled         & 23 & 15 & 16 & 15 & 7  & 0  & 416 & 354 \\
Neuroscience    & 63 & 12 & 9  & 1  & 54 & 11 & 331 & 317 \\
Origami         & 16 & 9  & 7  & 9  & 9  & 0  & 442 & 355 \\
\bottomrule
\end{tabular*}
\end{table}

To further characterize the qualitative difference between conditions, our professional describer consultant independently reviewed sample drafts using a hierarchical 0–5 rubric assessing whether each description is true, useful, focused, polished, and well-timed. GenAD drafts, while imperfect, had most tracks reaching at least the usefulness threshold and were correctable in place. Baseline drafts, by contrast, contained too many clips with extensive overlap between tracks, making it impractical to evaluate individual descriptions. The consultant noted that Baseline drafts, particularly for the neuroscience video, would likely require deleting all tracks and starting fresh rather than editing incrementally.

The quantitative differences and the professional describer's review indicate that participants encountered meaningfully different starting materials across the two AI conditions.

\subsection{Participants}
We recruited 30 participants (14 female, 16 male, ages 22–35) from a university campus. None had prior AD authoring experience, and about half (14) had never interacted with BLV users. This sample approximates the novice volunteer pool typical of crowd-sourced AD platforms but may not reflect the full diversity of that community in age, background, or ongoing motivation to contribute. Participants were compensated \$150 for their time.

\subsection{Procedure}
Prior to the session, participants received introductory materials including audio description guidelines and two 30-minute recordings of a professional describer authoring descriptions both from scratch and by editing an AI-generated draft in RefineAD.

The in-person session consisted of four phases: (1) an orientation covering the study goals, the RefineAD interface, and participant questions; (2) AD authoring for five videos, each assigned to one of the three conditions; (3) a NASA-TLX survey after each video to capture subjective workload; and (4) a post-session survey on overall experience. To keep sessions to a manageable length, participants used AI text-to-speech for voice delivery rather than recording their own voiceover.

Each participant completed one video from scratch, two with Baseline drafts, and two with GenAD drafts. The from-scratch condition was limited to low-difficulty cooking videos for two reasons. First, prior work established that novice describers require significant time to complete these videos from scratch, making higher-difficulty content impractical within a single session \cite{Yuksel2020Human-in-the-LoopUsers}. Extending this to high-difficulty content would have made sessions impractically long and introduced fatigue that could confound subsequent tasks. Second, using the same stimuli as \cite{Yuksel2020Human-in-the-LoopUsers} provides a behavioral reference point for comparing our from-scratch times against previous findings despite differences in interfaces and participant pools.

Participants were divided into four groups using a partial Latin square design. Condition assignments were counterbalanced so that each cooking video alternated between from-scratch and AI-assisted conditions across groups, while Videos 3–5 alternated between Baseline and GenAD across group pairs. Video order was randomized within each group to reduce order effects, and participants were not informed of the specific AI condition whenever possible (Appendix~D). Each video was completed without interruption to ensure accurate time measurement; participants could rest between tasks after completing the post-video survey.

\subsection{Measures}
We measured task completion time and assessed perceived workload using the NASA-TLX \cite{hart1988nasa} across conditions. Two-way repeated measures MANOVA was used to assess main effects and interactions between conditions with two levels of AI (Baseline and GenAD) and two difficulty levels (Low and High).  For low difficulty videos where time and perceived workload were assessed under all three conditions (From-scratch, Baseline, and GenAD), one-way repeated measures MANOVA was used to analyze differences between the conditions. Sample sizes vary slightly across comparisons due to missing timestamps or survey responses. Open-ended responses from post-study surveys were analyzed using inductive/deductive thematic consensus coding \cite{bradley2007qualitative}; two researchers independently coded all responses and reconciled discrepancies through discussion.

\subsection{Edit Log Analysis}
\label{sec:edit-taxonomy}
To characterize how participants modified the AI-generated drafts, we analyzed five dimensions of editing: deletions, insertions, text revisions, playback-type changes, and start-time adjustments. We first manually examined a randomized 45\% sample within each edit dimension (60 of 132 unique deletions, 53 of 118 insertions, 206 of 458 text revisions, 64 of 143 playback-type changes, and 78 of 174 start-time adjustments). For each sampled clip, we compared the AI-generated draft with the participant's final edit and, where relevant, the corresponding video and audio context. Based on recurring edit patterns, audio description guidelines, prior corpus studies, and feedback from accessibility consultants, we developed the following taxonomy:

\begin{enumerate}
\item \textbf{Deletion rationale.} Deleted clips were classified as \textit{incorrect} when the description was inaccurate or unsupported by the video~\cite{li2025vidhalluc}; \textit{redundant with audio} when the same information was already conveyed by dialogue, narration, or other soundtrack elements; \textit{redundant with AD} when nearby description clips already conveyed the information~\cite{itc2000}; or \textit{not useful} when the description did not provide useful visual information or was too vague, awkward, or incomplete to communicate it effectively~\cite{2024DescribedDCMP, remael2015adlab}.

\item \textbf{Insertion content.} Newly inserted information was classified as \textit{actions and expressions} (actions, movements, gestures, or facial expressions), \textit{people identity} (people, roles, or identifying information), \textit{objects and attributes} (objects and their visual properties), \textit{scene setting} (location, environment, spatial relationships, or temporal context), or \textit{on-screen text and data} (visible text, numbers, charts, or diagrams)~\cite{salway2007, piety2004, netflixad, 2024DescribedDCMP}. Multiple labels were assigned when an insertion added more than one type of information.

\item \textbf{Text revision.} Matched clips were classified as \textit{no change} when the text was retained verbatim; \textit{minor rewrite} when the wording changed without substantially changing the information conveyed; \textit{expanded} when information was added while retaining the original content; \textit{shortened} when information was removed while preserving the core description; or \textit{major rewrite} when the revision substantially changed the information conveyed~\cite{itc2000, snyder2014visual}.

\item \textbf{Playback adjustment.} Changes between inline and extended playback were classified according to the audio constraint they addressed. Extended-to-inline changes were classified as \textit{removed an unnecessary pause} when the description could fit within the available audio gap without requiring extended playback. Inline-to-extended changes were classified as \textit{avoided dialogue overlap} when extended playback prevented the description from overlapping with speech~\cite{wcag21}.

\item \textbf{Timing adjustment.} For matched clips whose start time shifted by more than $0.5\,\mathrm{s}$, changes were classified as \textit{avoided dialogue overlap}, \textit{accommodated text revision}, or \textit{resynchronized with visual event}, depending on whether the new timing better accommodated speech, the revised description length, or the corresponding visual event~\cite{itc2000}.
\end{enumerate}

We then applied the taxonomy to the full dataset using task-specific
model-assisted classification and deterministic rules. GPT-5.6 Sol was
used for semantic and visual judgments, with the evidence provided to
the model tailored to each edit type. For deletions, the model received the deleted description together with timestamped video frames sampled before, during, and after the clip, the surrounding transcript or audio context, and nearby descriptions retained in the participant's final track. It followed a fixed decision order to assess whether the deleted content was unsupported by the video, redundant with retained AD, redundant with the soundtrack, or not useful. For insertions, the model compared each newly inserted description against the complete original AI draft and labeled only information that was semantically novel, allowing multiple content categories when applicable. For text revisions, the model compared the original AI description with the participant's final revision and classified whether information was preserved, added, removed, or substantially changed.

Model outputs were constrained to structured JSON schemas containing
the predefined labels together with task-specific evidence fields.
Programmatic validation checked label--evidence consistency and retried responses that violated the expected schema. Full prompts, output schemas, and classification code are provided in
Appendix~\ref{appendix:classification-prompts} and the supplementary
materials.

Playback adjustments were classified deterministically using edit
metadata, actual description duration, dialogue intervals, and overlap
with other retained descriptions. Timing adjustments followed a
hierarchical procedure: changes that reduced dialogue overlap or
accommodated revised description duration were identified
deterministically; for the remaining cases, GPT-5.6 Sol compared
timestamped frames around the original and edited positions to determine whether the shift better synchronized the description with the visual event. Changes unsupported by these criteria were retained as \textit{Other} rather than assigned a rationale.

All model-generated labels were independently reviewed by two
researchers against the corresponding AI draft, participant edit, and
relevant video/audio context. Each reviewer either confirmed the
model-assigned category or proposed a different classification. Cases
in which either reviewer disagreed with the model label, or the two
reviewers assigned different categories, were discussed and reconciled. The reconciled human-reviewed labels were used in all subsequent analyses.

\section{Results}

\subsection{Authoring Efficiency and User Experience}
\label{sec:efficiency}

\subsubsection{Task Completion Time}
\label{sec:task-time}

\begin{figure}[!htb]
  \centering
  \includegraphics[width=0.50\textwidth]{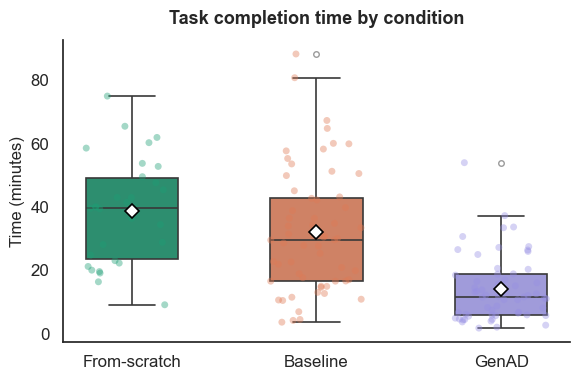}
  \caption{Task completion time by condition. GenAD ($M = 13.9$ min) was significantly faster than Baseline ($M = 31.0$ min) and From-scratch ($M = 38.4$ min). Diamonds indicate means.}
  \label{fig:time}
\end{figure}

A two-way repeated measures MANOVA revealed a significant main effect for AI condition on task completion time (Pillai's trace $= 0.587$, $p < 0.001$), with no significant main effect of video difficulty (Pillai's trace $= 0.29$, $p = 0.396$) or AI condition $\times$ video difficulty interaction (Pillai's trace $< 0.01$, $p = 0.985$). Participants using GenAD completed descriptions in approximately 14 minutes for both low-difficulty ($M = 13.98$, $SD = 9.94$) and high-difficulty ($M = 13.85$, $SD = 11.22$) videos, compared with roughly 30 minutes for Baseline ($M = 29.99$, $SD = 18.16$; $M = 31.92$, $SD = 19.23$). For low-difficulty videos, where all three conditions were available, from-scratch description ($M = 38.44$, $SD = 16.75$) took significantly longer than Baseline editing ($t(27) = 1.93$, $p = 0.032$), which took significantly longer than GenAD editing ($t(28) = 2.02$, $p = 0.027$). Figure~\ref{fig:time} shows the distribution of completion times across conditions.

\subsubsection{Perceived Workload}
The two-way repeated measures MANOVA found a significant main effect of AI condition on perceived workload (Pillai's trace $= 0.135$, $p = 0.046$), with no significant effect of video difficulty or interaction. GenAD was rated as less demanding than Baseline for both easy ($M = 29.22$ vs.\ $37.24$) and hard ($M = 33.36$ vs.\ $45.37$) videos. Among easy videos, where all three conditions were available, only the difference between from-scratch ($M = 42.36$) and GenAD was significant ($t(28) = 2.41$, $p = 0.012$). Table~\ref{tab:tlx} reports workload scores by condition and difficulty.

\begin{table}[!htb]
  \centering
  \caption{NASA-TLX overall workload by condition and video difficulty (0--100 scale; higher = greater demand).}
  \label{tab:tlx}
  \small
  \begin{tabular}{lcccc}
    \toprule
    & \multicolumn{2}{c}{\textbf{Low Difficulty videos}} & \multicolumn{2}{c}{\textbf{High Difficulty videos}} \\
    \cmidrule(lr){2-3} \cmidrule(lr){4-5}
    \textbf{Condition} & $M$ & $SD$ & $M$ & $SD$ \\
    \midrule
    From-scratch & 42.36 & 23.79 & ---   & ---   \\
    Baseline     & 37.24 & 25.21 & 45.37 & 26.97 \\
    GenAD        & \textbf{29.22} & \textbf{20.26} & \textbf{33.36} & \textbf{19.96} \\
    \bottomrule
  \end{tabular}

  \par\vspace{4pt}
  {\footnotesize\raggedright\textit{Note.} Bold indicates the lowest workload per difficulty level.\par}
\end{table}

\subsubsection{System Preference and Perceived Speed}
Of the 30 participants, 23 (76.7\%) preferred the GenAD-RefineAD system, while 5 (16.7\%) preferred authoring from scratch and 2 (6.7\%) had no preference. Similarly, 23 (76.7\%) perceived the system as quicker, 4 (13.3\%) found from-scratch faster, and 3 (10.0\%) saw no difference. Preference and perceived speed aligned for 24 of 30 participants.

\begin{figure}[!htb]
  \centering
  \includegraphics[width=0.35\linewidth]{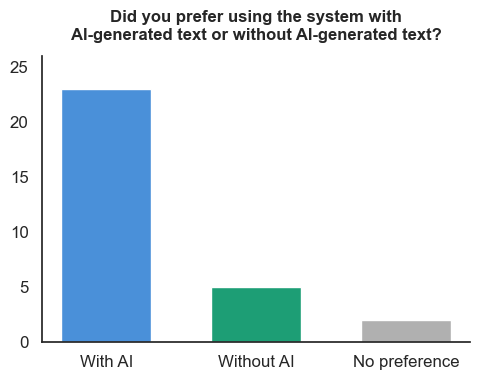}%
  \hspace{0.04\linewidth}%
  \includegraphics[width=0.35\linewidth]{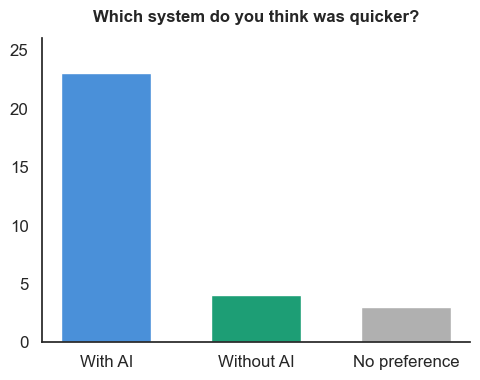}
  \caption{Post-study preference responses ($N=30$). Left: overall preference for authoring with vs.\ without AI-generated text. Right: perceived speed of each system.}
  \label{fig:preferences}
\end{figure}

\FloatBarrier

\subsection{How Participants Edited AI Drafts}
\label{sec:editing-drafts}

\subsubsection{Editing Feature Usage}
\begin{figure}[!htb]
  \centering
  \includegraphics[width=0.50\textwidth]{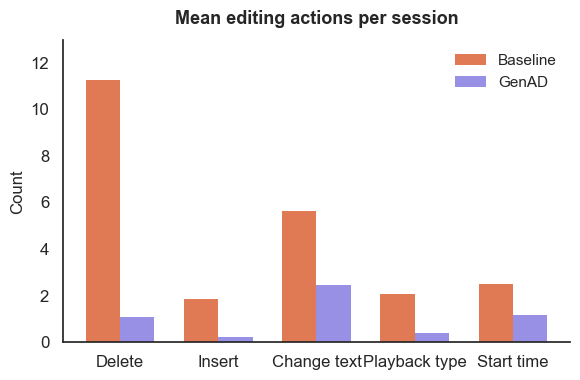}
  \caption{Mean number of editing actions per session by condition.}
  \label{fig:editing}
\end{figure}

The edit-log analysis included 112 AI–human collaboration sessions: 57 in the Baseline condition and 55 in the GenAD condition. We analyzed how participants used RefineAD's editing features across the Baseline and GenAD conditions. Figure~\ref{fig:editing} shows the mean number of editing actions per session. Participants made fewer edits in the GenAD condition across all features. The largest difference was in deletions: Baseline descriptions averaged 11.3 deletes per session compared to 1.1 for GenAD, indicating that the unguided AI system produced excess or irrelevant clips that participants needed to remove. Text edits followed a similar pattern (5.6 vs.\ 2.4), as did playback type changes (2.1 vs.\ 0.4) and start time adjustments (2.5 vs.\ 1.2). Insertions were infrequent in both conditions (1.9 vs.\ 0.2). Overall, participants made fewer edits across all feature types in the GenAD condition, consistent with higher initial draft quality.

\subsubsection{Editorial Functions}
\begin{figure*}[!htb]
  \centering
  \includegraphics[width=\textwidth]{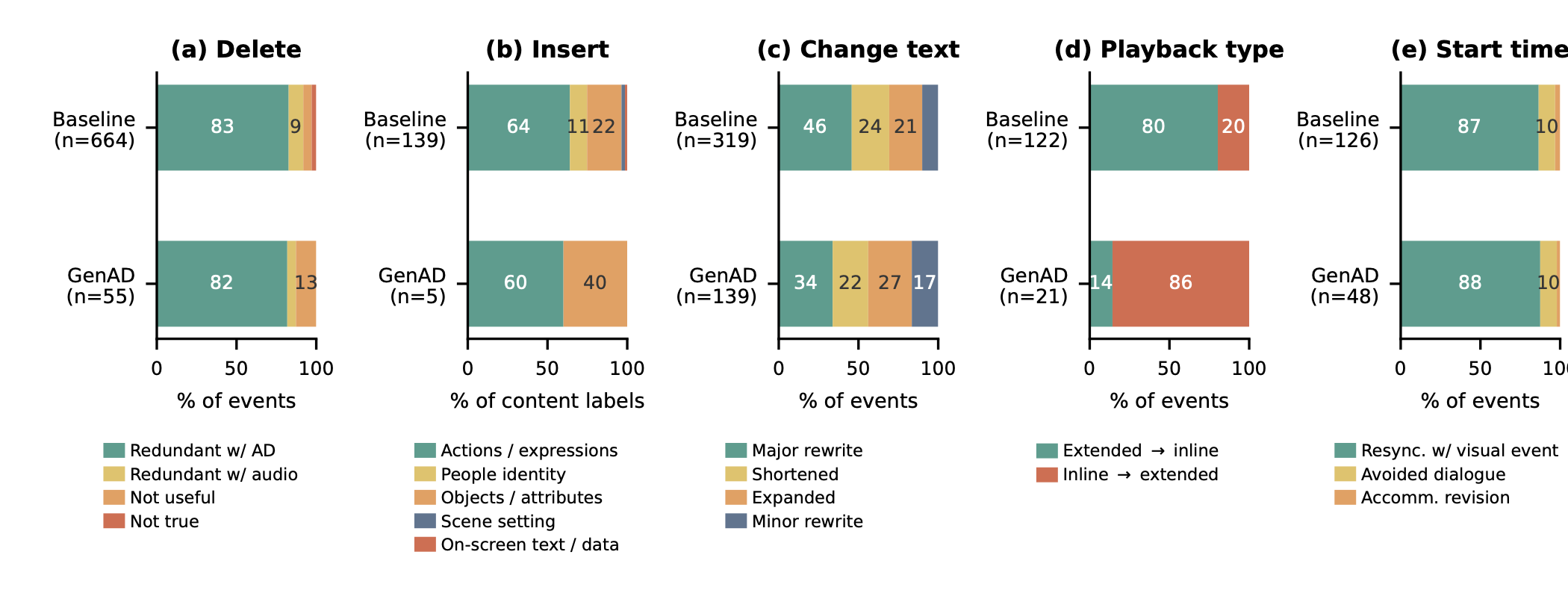}
  \caption{Classification of edits by condition: (a) deletion rationales, (b) insertion content types, shown as a share of content labels because an insertion can carry multiple labels, (c) text-revision types, (d) direction of playback changes, and (e) start-time rationales. All bars show within-condition percentages; GenAD denominators are small.}
  \label{fig:edit-composition}
\end{figure*}

Each change was assigned an editorial function using the taxonomy in Section~\ref{sec:edit-taxonomy}. Figure~\ref{fig:edit-composition} shows how edits were distributed across categories in each condition. Appendix~\ref{appendix:examples} lists representative examples of each.

\paragraph{Deletions.}
Deletions overwhelmingly targeted redundant content. Of the 664 Baseline and 55 GenAD deletions, redundancy with nearby description clips accounted for the large majority in both conditions (Baseline: 549 of 664, 83\%; GenAD: 45 of 55, 82\%; Figure~\ref{fig:edit-composition}a). The remaining categories were small: redundancy with the audio track (Baseline: 61, 9\%; GenAD: 3, 5\%), \textit{not useful} content (Baseline: 35, 5\%; GenAD: 7, 13\%), and \textit{not true} descriptions (Baseline: 19, 3\%; GenAD: 0). Deletion activity also concentrated sharply by video. The Neuroscience Baseline video, whose soundtrack carries continuous narration, alone accounted for 423 of the 664 Baseline deletions (64\%) and for 61 of the 64 audio-redundancy deletions (95\%) in the dataset. All 19 \textit{not true} deletions occurred on the Origami Baseline video, the more difficult of the instructional videos.

\paragraph{Insertions.}
Most Baseline insertions (95 of 108, 88\%) introduced information absent from the AI draft. Most inserted content described actions and expressions, appearing in 89 of 108 insertions and accounting for 64\% of all content labels. Objects and attributes accounted for 22\% of labels, while people identity accounted for 11\%. The remaining two categories together accounted for 3\% (Figure~\ref{fig:edit-composition}b).
Insertions concentrated on the \textit{Tangled} video: 93 of the 108 Baseline insertions (86\%) occurred here, and GenAD received 10 insertions from 6 participants; 3 of these insertions introduced novel information.

\paragraph{Text revisions.}
Text revision was the most frequent editing action on GenAD drafts, where it outnumbered deletion, and it drew the broadest GenAD participation, with 22 participants revising at least one clip. Major rewrites were the most common revision type in both conditions (Baseline: 46\%; GenAD: 34\%), with shortenings, expansions, and minor rewrites splitting the remainder; GenAD revisions skewed proportionally toward expansions and minor rewrites (Figure~\ref{fig:edit-composition}c). Unlike deletions and insertions, revisions spread across all five videos in both conditions.

\paragraph{Playback changes.}
Playback changes ran in opposite directions across conditions. Baseline editors predominantly converted extended clips to inline clips (98 of 122, 80\%), and 86 of these were classified as removing an unnecessary pause: the description fit within the available audio gap without pausing the video (Figure~\ref{fig:edit-composition}d). GenAD editors predominantly converted inline clips to extended ones (18 of 21, 86\%), and 15 of these were classified as avoiding overlap with speech or another description. Baseline playback changes concentrated on the Origami and Tangled clips (95 of 122, 78\%), while the deletion-heavy Neuroscience clip drew almost none.

\paragraph{Start-time adjustments.}
Baseline start-time shifts were larger and more variable than GenAD shifts (median $|$shift$|$ = 3.2\,s vs.\ 2.4\,s; maximum 93.4\,s vs.\ 44.6\,s), and nearly all Baseline shifts (94\%) accompanied a change to the clip's text, compared with about half of GenAD shifts. Resynchronizing the description with the visual event it covers accounted for the large majority of shifts in both conditions (Baseline: 109 of 126, 87\%; GenAD: 42 of 48, 88\%), with overlap avoidance and text accommodation infrequent (Figure~\ref{fig:edit-composition}e). Start-time adjustments were also the most evenly spread action across videos.

\paragraph{Other edits.}
A small share of edits fell outside the defined editorial functions. In total, 20 insertions introduced no new information: 11 reworded an existing clip for style, clarity, or word choice, but did so by inserting a new clip and deleting the old one instead of revising it in place, and 9 relocated existing text to a new position. In addition, 15 playback changes did not correspond to any audio constraint. Most of these (13 of 15) were inline-to-extended conversions: the editing interface assigns extended playback by default after a text edit, and describers did not always revert these clips to inline even when the revised description fit the available gap. Together these edits amounted to about 2\% of all edit events.
\FloatBarrier

\subsection{What Participants Kept Unchanged}
\label{sec:preserved}

Across both conditions, 1,663 untouched instances corresponded to 279 unique draft clips (99 GenAD, 180 Baseline). Because the number of participants differed across videos, we calculated each clip's acceptance rate as the percentage of participants assigned to that video who left it completely unchanged. Table~\ref{tab:preserved-clips} summarizes the degree of agreement across editors; acceptance was markedly higher for GenAD clips (median 80.0\%) than for Baseline clips (median 57.1\%).

\begin{table}[!htb]
\centering
\caption{Acceptance of preserved draft clips across participants.}
\label{tab:preserved-clips}
\small
\renewcommand{\arraystretch}{1.3}
\begin{tabular*}{\columnwidth}{@{\extracolsep{\fill}}lr}
\toprule
\textbf{Measure} & \textbf{Result} \\
\midrule
Unique preserved clips & 279 \\
Median acceptance rate & 64.3\% \\
Accepted by $\geq$50\% of participants & 73.5\% \\
Accepted by $\geq$70\% of participants & 45.9\% \\
\bottomrule
\end{tabular*}
\end{table}

The most widely preserved clips shared three characteristics. First, they described a single, discrete visual event with a concrete action: \textit{``Rapunzel wedges a green chair against the door handle''} (GenAD \textit{Tangled}) was preserved by all 14 participants. Second, they captured clearly verifiable on-screen information, particularly meaningful labels in educational content; a description identifying the \textit{``AXON HILLOCK''} and its location on the neuron diagram (GenAD \textit{Neuroscience}) was accepted by 13 of 15 participants. Third, they described stable visual states or objects: \textit{``A frying pan filled with hot oil; several potato balls are frying, surrounded by bubbling oil''} (Baseline \textit{Potato Balls}) was preserved by six of seven participants. Among unique preserved GenAD clips, 83.8\% used inline playback.

\subsection{Contribution Score}

Table~\ref{tab:contribution-scores} summarizes AI and human contribution by condition. GenAD drafts were largely retained ($M = 93\%$, $SD = 9\%$), with volunteers contributing only 7\% on average. Baseline drafts required substantially more editing ($M = 67\%$, $SD = 28\%$) with wider variance across volunteers.

\begin{table}[!htb]
\centering
\caption{AI attribution by condition.}
\label{tab:contribution-scores}
\small
\begin{tabular*}{\columnwidth}{@{\extracolsep{\fill}}llcccc}
\toprule
\textbf{Condition} & \textbf{n} & \textbf{AI Attribution} & \textbf{Human Contribution} \\
\midrule
Baseline   &  57 & $67\% \pm 28\%$ & $33\% \pm 28\%$ \\
GenAD      &  55 & $\mathbf{93\% \pm 9\%}$  & $\mathbf{7\% \pm 9\%}$   \\
\bottomrule
\end{tabular*}
\par\vspace{4pt}
{\footnotesize\raggedright\textit{Note.} Bold indicates the highest AI retention (lowest human editing effort).\par}
\end{table}

\begin{figure}[!htb]
  \centering
  \includegraphics[width=\columnwidth]{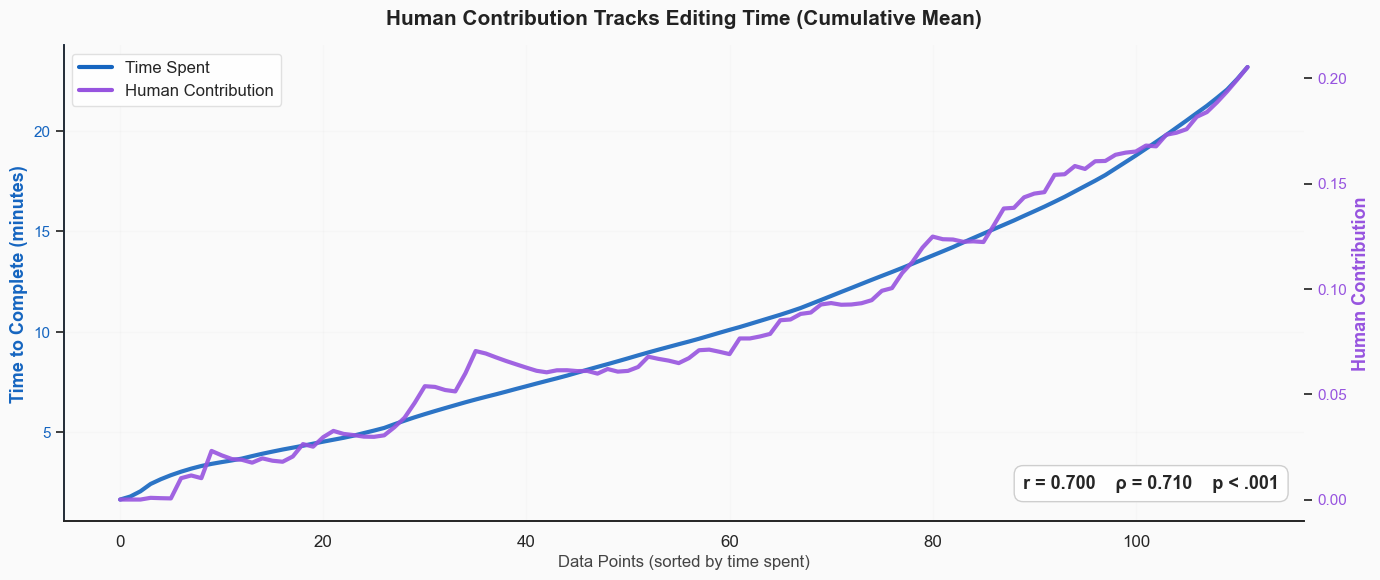}
  \caption{Human contribution tracks task completion time across participants ($\rho = .71$, $p < .001$), indicating the score's alignment with observed editing behavior.}
  \label{fig:contribution-trend}
\end{figure}

As an initial validity check, we correlated it with task completion time (Section~\ref{sec:task-time}) and compared it against normalized Levenshtein distance over the concatenated clip texts. Both the MDCI and the Levenshtein method showed strong positive correlations with completion time ($\rho = .71$ and $.68$, $p < .001$). Across all conditions, MDCI yielded consistently stronger correlations than Levenshtein. While the correlation was lower in the GenAD condition ($r = .40$ vs.\ $\rho = .59$), this was primarily due to the low variance in those scores ($SD = 9\%$); nevertheless, rank order associations remained moderate. As shown in Figure~\ref{fig:contribution-trend}, human contribution tracks closely with the time spent by each participant.

\begin{table}[!htb]
\centering
\caption{Correlation between human contribution and task completion
time by condition.
$^{*}p < .01$, $^{**}p < .001$.}
\label{tab:contribution-correlation}
\small
\begin{tabular*}{\columnwidth}{@{\extracolsep{\fill}}llcccc}
\toprule
 & & \multicolumn{2}{c}{\textbf{MDCI}} & \multicolumn{2}{c}{\textbf{Levenshtein}} \\
\cmidrule(lr){3-4} \cmidrule(lr){5-6}
\textbf{Condition} & \textbf{n} & \textbf{Pearson} & \textbf{Spearman} & \textbf{Pearson} & \textbf{Spearman} \\
\midrule
Overall    & 112 & $.70^{**}$ & $.71^{**}$ & $.66^{**}$ & $.68^{**}$ \\
Baseline   &  57 & $.63^{**}$ & $.63^{**}$ & $.57^{**}$ & $.59^{**}$ \\
GenAD      &  55 & $.40^{*}$  & $.59^{**}$ & $.35^{*}$  & $.53^{**}$ \\
\bottomrule
\end{tabular*}
\end{table}

\FloatBarrier

\subsection{Open-Ended Feedback}
\begin{figure*}[t]
  \centering
  \includegraphics[width=\textwidth]{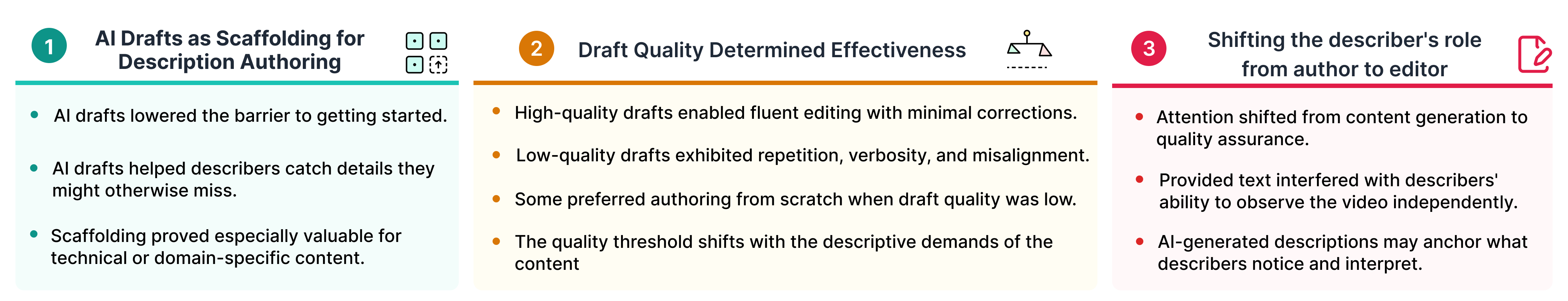}
  \caption{Three themes from participants' qualitative responses: (1) AI drafts as scaffolding for description authoring, (2) Draft quality determined effectiveness, and (3) AI drafts shifted the describer's role from author to editor, with both benefits and drawbacks for the editing process.}
  \label{fig:theme}
\end{figure*}

Three themes emerged from participants' open-ended responses about their experience across conditions (Figure~\ref{fig:theme}).

\subsubsection{AI Drafts as Scaffolding for Description Authoring}

Participants frequently described AI drafts as a useful starting point ($n=16$). Rather than generating descriptions from a blank page, they could focus on checking and refining existing content: \textit{``Having an initial draft helped me focus on accuracy, timing, and clarity rather than starting from a blank page.''} Drafts also helped participants notice visual information they might otherwise miss ($n=8$), with one participant noting that \textit{``you don't end up missing things in the video.''}

This support was particularly valuable for visually complex or technical content. Participants described the neuroscience video as difficult to author from scratch because it required explaining diagrams and specialized visual information. In these cases, the draft provided a foundation for content that novice describers found difficult to formulate independently. Clips conveying concrete actions, visual states, and diagram information were among those most often preserved (Section~\ref{sec:preserved}).

\subsubsection{Draft Quality Determined Effectiveness}

The usefulness of AI assistance depended strongly on draft quality ($n=13$). High-quality drafts enabled participants to make relatively minor refinements ($n=8$): \textit{``When the suggested text is good quality, it only requires minor edits.''} In contrast, low-quality drafts could create more work than authoring from scratch ($n=4$): \textit{``I spent quite a lot [of] time correcting it instead of having the time to describe it.''}

Participants most often attributed this burden to repetition, unnecessary detail, and incorrect timing. These concerns parallel the editing patterns observed above, where participants deleted redundant content and adjusted text, playback, and timing. The effect was also content-dependent. Lower-quality drafts were especially frustrating for technical or spatially complex videos such as neuroscience and origami, whereas participants raised fewer concerns for more familiar sequential content such as cooking. 

\subsubsection{From Authoring to Editing}

AI drafts also changed the nature of the task. With a description already provided, participants described focusing on verification and refinement rather than deciding what to describe from scratch: \textit{``If the provided text is in a high quality, it would be easier to check the accuracy.''} This shift is consistent with the high retention of GenAD drafts and the relatively limited human contribution observed in the quantitative results.

However, several participants also described drawbacks to this editorial mode. Existing text could compete with attention to the video, create cognitive overload, or influence what participants noticed. One participant described the text as \textit{``interrupting watching the video,''} while another preferred working without it to avoid \textit{``bias for observation.''} Although AI drafts lowered the effort required to begin authoring, they could also direct attention toward the system's framing of the visual content.

\section{Discussion}

Novice describers substantially reshaped AI-generated descriptions, but their edits did not always address problems in the initial drafts. We examine how these patterns varied with draft quality and the different editorial demands it imposed.

\subsection{How Novice Describers Reshape AI Drafts}

Edits extended beyond local wording changes to include reorganizing clips, adding missing information, adjusting timing, and adapting descriptions to better fit the structure and tone of the video. Together, these changes show that novice describers were not passive reviewers of AI output, but actively reshaped the generated drafts.

\subsubsection{Restructuring clip boundaries.}
\label{sec6:restructing}
Editors reshaped how information was divided across clips in both directions. In the \textit{Potato Balls} cooking video, related information was sometimes spread across several short clips. A visual action and its accompanying on-screen text, for example, could appear separately. Editors combined these into a single description, such as merging a visual description of adding butter with the on-screen text \textit{``...and black pepper! The best spice of the entire world!''} Because the system uses different voices for visual description and on-screen text, adjacent clips also created frequent voice switching, especially in text-dense videos. Although this distinction was designed to avoid repeatedly announcing \textit{``text on screen''} (Section~3.1.4), participants found frequent switching distracting, a concern also raised by our blind QC consultant. Combining the clips therefore addressed a problem that arose from the interaction between the system's clip segmentation and voice design.

Editors also split longer descriptions when the described actions occurred at different times. In \textit{Tangled}, the description \textit{``Rapunzel moves slightly closer, her eyes widening. Flynn suddenly opens his eyes''} was shortened, with \textit{``Flynn suddenly opens his eyes''} moved into a separate clip at the moment it occurred.

\subsubsection{Adding missing information}

Participants also filled gaps in the drafts by adding omitted actions and instructional details. In the \textit{Tangled} video, for example, editors inserted actions such as \textit{``He slips through the window and shuts it behind him''} and \textit{``With a smile, he opens his bag to check inside.''} These additions supplied events that were visually important but absent from the generated descriptions.

The \textit{Crispy Fritters} video further illustrates how draft quality affected the recovery of missing information. The Baseline draft omitted the displayed instruction \textit{``1/2 tbs oil,''} which was already captured by GenAD; 3 of 6 participants working from the Baseline draft added this detail. Both drafts also missed the prominently displayed instruction \textit{``1/2 tsp baking powder.''} This omission was recovered by 5 of 8 participants in GenAD, compared with only 1 of 6 in Baseline, while 14 of 16 participants working From Scratch included it. Because participants in both GenAD and From Scratch frequently captured the instruction, the information itself appears relatively salient. The lower recovery in Baseline occurred alongside substantially more correction activity, suggesting that even salient omissions can remain unaddressed when the starting draft requires extensive revision.

\subsubsection{Adapting Style and Tone}

In the cooking and educational videos, describers often reframed descriptions as direct instructions, making them better fit the procedural nature of the content. In \textit{Potato Balls}, \textit{``A hand places a potato ball into hot oil in a frying pan. The oil bubbles around the potato ball, indicating it's frying''} was rewritten as \textit{``Put enough olive oil in a pan. Heat the oil until it makes the sound of bubbles. Then put the potato balls one by one into the boiling oil to fry them.''} In \textit{Crispy Fritters}, \textit{``Oil is added to the mixture, continue to mix till fully incorporated''} became \textit{``Add 1/2 tbs oil to the mixture, continue to mix till fully incorporated,''} replacing passive phrasing with a direct instruction while also adding the exact measurement. This shift toward active language is consistent with established audio description guidelines \cite{snyder2014visual}. 

Similar adaptations appeared in other instructional videos. In the \textit{Origami} tutorial, one participant added, \textit{``This is a paper folding tutorial. Using just one piece of paper, you can fold an origami fish,''} orienting the viewer to the task before the individual steps began. These edits show that novice describers reframed descriptions to better fit the purpose and presentation of the video. They also introduced stylistic choices and individual expression that were less evident in the AI-generated drafts.

\subsection{Challenges in Editing AI Drafts}
Some problems in the initial drafts remained unresolved after revision, while other edits introduced new issues. The following cases show that reducing the burden of writing does not remove the need for careful review. A usable draft may help novices recover obvious omissions, while subtle errors or unsupported details can still remain unnoticed or be introduced during editing. 

\subsubsection{Draft Wording Can Anchor Human Edits}
\label{sec6:anchor}
The presence of human editing did not ensure that problems in the draft would be identified or corrected. Novice repairs could coexist with uncorrected problems in the same description. In \textit{Tangled}, for example, a participant added the missing hiding action \textit{``The woman is hiding behind the clothes hanger,''} improving visual coverage, but continued referring to \textit{Rapunzel} as \textit{``the woman,''} following the terminology established by the Baseline draft. Character identification was a known weakness of that generation approach (Section~3.1.5). This suggests that AI wording could anchor not only what novices describe but also how they describe it, consistent with prior work on anchoring in AI-assisted writing~\cite{jakesch2023cowriting, agarwal2025homogenize}.

\subsubsection{Subtle Inaccuracies Can Remain Unnoticed}
None of the deleted GenAD clips were classified as not true (Figure~\ref{fig:edit-composition}a). However, this does not mean that GenAD contained no inaccuracies. Closer inspection showed that some subtle errors were retained by most editors. Incorrect details were harder to spot when they involved subtle spatial errors. In the \textit{Origami} video, participants corrected a description stating that the paper was folded diagonally when it was actually folded horizontally. However, only one of 13 editors corrected an instruction to \textit{``turn the paper 45 degrees''} when the video showed a 90-degree rotation. 

A similar pattern appeared in \textit{Neuroscience}. The generated description stated that \textit{``a circle in black ink is drawn around the yellow center of the neuron,''} although the circle is drawn around the center of the dendritic section rather than the center of the neuron as a whole. Our professional describer flagged this error and instead described it as \textit{``a circle in black ink is drawn around the \textbf{root ball} of the neuron.''} Only two of 15 editors revised this description: one removed \textit{``center''} and one specifically revised to \textit{``center of the yellow body of the neuron.''} This contrasts with \textit{Crispy Fritters}, where most GenAD editors recovered a conspicuous missing instruction. Together, the cases suggest that editing performance may depend not only on whether information is missing or incorrect, but also on how perceptually salient and easy to verify that information is.

\subsubsection{What Novice Describers May Overlook}

Expanded descriptions could also introduce information that was not directly supported by the video. In \textit{Crispy Fritters}, one participant rewrote \textit{``Golden fritters with chili sauce on a white plate''} as \textit{``In the end we have these golden crispy fritters which can be served with ketchup.''} The video showed only a red sauce, with no on-screen text or soundtrack identifying it. The original \textit{``chili sauce''} detail came from the video's YouTube description, which was included as metadata in the generation pipeline but was not shown in the editing interface. In this case, an edit intended to improve the description unintentionally introduced an unverifiable detail.

Some inserted descriptions also moved beyond directly observable information. In the \textit{Tangled} video, for example, participants added interpretations of characters' mental states, such as \textit{``afraid that he may wake up at any moment.''} Similarly, some rewrites adopted conversational tutorial language (e.g., \textit{``we use''} and \textit{``in the end''}), shifting the description away from neutral visual reporting toward a more interpretive register. Whether such changes constitute improvements remains contested. Traditional audio description guidelines generally position interpretation as a departure from the describer's role~\cite{snyder2014visual, fryer2016introduction}, whereas more recent work suggests that interpretive description may better support BLV audiences in some contexts~\cite{walczak2017creative, schaefferlacroix2023beyond}.

\subsection{Adapting to Content and Describer Needs}

Our quantitative results show that AI assistance was substantially more useful when the generated draft was higher quality. On low-difficulty videos, moving from From Scratch to Baseline reduced completion time by about 8 minutes and workload by 5 points, whereas moving from Baseline to GenAD produced a larger reduction of about 16 minutes and 8 points. This pattern is consistent with prior work showing that the usefulness of AI-assisted writing depends strongly on the quality of the provided suggestions \cite{Green2013TheTranslation, daems2017identifying, dhillon2024shaping}. Our editing results further suggest that draft quality alone does not determine how useful that assistance will be. Editors could readily improve drafts when the relevant information was visually salient, but were less likely to catch subtle spatial or technical inaccuracies. A strong draft can therefore reduce the work of composing and restructuring descriptions while still leaving a demanding verification task when correctness requires precise visual or domain-specific judgment.

This distinction has implications for AI-assisted AD systems. For visually explicit or procedural content, a strong initial draft may provide an effective scaffold for editing. Content involving fine spatial relationships or specialized terminology may require additional context or support for verifying generated descriptions rather than simply more fluent drafts. Because we did not observe a significant AI condition $\times$ video difficulty interaction, we treat these content-dependent differences as directions suggested by our qualitative and editing evidence rather than a confirmed quantitative effect.

Generation strategies should adapt to the specific demands of the content. For instance, visually concrete or sequential videos may require only lightweight prompting, whereas technical or spatially complex material may benefit from additional context and verification support. The text-heavy videos provide another example: a voice design intended to distinguish visual descriptions from on-screen text became distracting when the text appeared frequently (Section~\ref{sec6:restructing}). For text-dense videos, merging the two into a single voice may therefore be preferable, while separate voices may remain useful when on-screen text appears only occasionally.

The editing interface should likewise adapt to content and user needs. Our consultants noted that experienced describers adjust priorities by genre: step-by-step accuracy for instructional content, and character and pacing for entertainment. Novices may not yet have these genre-specific strategies, so surfacing a video's genre or communicative goal could help guide editing priorities. The interface could also accommodate individual preferences. For example, several non-native English speakers preferred simpler vocabulary, while describers varied in whether they preferred to start from more detailed or more concise descriptions. Although we did not formally evaluate these customization options, informal feedback from participants suggests that allowing users to adjust language complexity and level of detail could better support different editing preferences.

A reliable and approachable editing workflow is also essential. Because AI-assisted authoring shifts much of the work toward editing, technical friction can directly interfere with that process. Participants reported needing to refresh the page for audio to play after edits, and some initially found the feature set overwhelming. In response, we improved playback reliability and added a guided tutorial mode (Appendix~\ref{appendix:tutorial}). Reducing such barriers may be particularly important for novices, whose early experience can shape continued participation~\cite{jamison2003turnover}.

Beyond usability, interfaces should also support critical evaluation of AI-generated drafts. Given the risk of anchoring on initial AI wording, supported by related work \cite{jakesch2023cowriting, agarwal2025homogenize} and our findings in Section~\ref{sec6:anchor}, systems could delay the draft until after an initial viewing, highlight portions that may require verification, or gradually reduce scaffolding as describers gain experience~\cite{lee2024designspace}.

\subsection{Measuring Contribution Through Editing Behavior}
Our edit-log analysis shows that describers reshaped drafts in several ways: they deleted and inserted clips, rewrote text, changed playback modes, and adjusted timing. Voice delivery represents another dimension of contribution, as vocal rendering can affect how the same description is experienced~\cite{Walczak2018VocalPresence, Fernandez-Torne2015TheCatalan}. Voice is therefore included in MDCI, although participants did not use this feature in the present experiment. These dimensions motivated the Multi-Dimensional Contribution Index (MDCI): the edit taxonomy captures \textit{what} describers change, while MDCI measures \textit{how much} those changes shape the final description.

Text similarity alone, as used in traditional edit-distance measures, can underestimate this contribution. A describer may preserve most of an AI-generated sentence while changing when it plays or whether the video pauses for it. These decisions affect the final listening experience while leaving little textual trace. Even within the text dimension, the number of words changed does not necessarily reflect the significance of an edit. MDCI therefore considers semantic and stylistic similarity alongside lexical distance. The instructional rewrites in Section~6.1 are one example: recasting a passive observation as a direct instruction alters the style of a description more than its wording. Similarly, in the \textit{Origami} video (Section~6.2.2), changing \textit{``turn the paper 45 degrees''} to \textit{``turn the paper 90 degrees''} changes only a single number but corrects an inaccurate spatial instruction. MDCI was strongly associated with task completion time and tracked time better than normalized Levenshtein distance. The margin is small, but it suggests that the added dimensions capture real editing work.

MDCI nevertheless remains an imperfect measure of editorial effort. Large contributions can reflect repetitive edits requiring little judgment, as in Baseline deletions of redundant clips, while small contributions can include careful repairs such as spatial corrections in GenAD. Similar mismatches have been observed in machine translation, where small edits may require substantial judgment and large edits may be largely mechanical~\cite{krings2001repairing, koponen2012comparing}. Promising directions include weighting edits by their taxonomy category, using process signals such as time spent or video replays before an edit as proxies for judgment, and calibrating scores against BLV audience assessments of the resulting descriptions.

Accurately capturing this contribution becomes especially important as AI-assisted authoring moves from individual use toward community-driven platforms. The difference in AI retention between GenAD ($M = 93\%$, $SD = 9\%$) and Baseline ($M = 67\%$, $SD = 28\%$) shows how strongly the balance of human and AI work can depend on the starting draft. Making that balance visible can give volunteers appropriate credit, clarify how much of a final description comes from AI versus human revision, and reveal where generated descriptions consistently require intervention. For novices, viewing how others edited the same draft functions as informal training. The opt-in design preserves authorial agency, maintaining the sense of ownership that motivates contribution \cite{kadoma2024ownership}.

MDCI itself also requires further validation. Its current weights ($w_{\text{text}} = w_{\text{time}} = 0.45$; $w_{\text{pb}} = w_{\text{voice}} = 0.05$) were established through expert consultation. Future work should examine how well these weights correspond to describers' perceived effort and ownership, and whether the relative importance of text, timing, playback, and delivery varies across content types or editing contexts.

\section{Limitations}
GenAD relies on GPT-4o, whose computational cost scales with video volume and whose output characteristics shape the editing patterns we observed. Different models may produce drafts with different error profiles, changing what editors need to remove or revise and how they interact with the text. Experimenting with a range of multimodal models, including smaller, open-source alternatives, would clarify whether these behaviors are model-specific or generalize across draft sources, while also reducing cost for deployment at scale. Our participant sample may also not fully reflect the diversity of the broader crowd-sourced AD community. This study focused on the authoring experience and did not evaluate the final descriptions. Prior work found that BLV users rated human-AI co-authored descriptions higher in quality and topic understanding than manually authored ones \cite{Yuksel2020Human-in-the-LoopUsers}; whether GenAD drafts yield similar or stronger outcomes across content types is an immediate next step.

\section{Conclusion}

In this work, we studied how novice describers edit AI-generated audio description drafts of differing quality. GenAD drafts cut completion time by more than half and significantly reduced workload, while unguided Baseline drafts offered only modest time savings over writing from scratch. GenAD drafts also required much less deletion and fewer major rewrites than Baseline drafts. Novice describers made substantial changes to content, timing, structure, and style, and MDCI captured these contributions where text-only measures would not. Higher-quality drafts did not eliminate the need for careful judgment: editors could miss subtle inaccuracies or anchor to wording in the initial draft. Together, these findings suggest that effective human-AI authoring depends on improving generated drafts and on designing workflows that support human judgment, refinement, and meaningful contribution.

\bibliographystyle{ACM-Reference-Format}
\bibliography{references}

@article{Pitcher-Cooper2023YouDataset,
    title = {{You Described, We Archived: A Rich Audio Description Dataset}},
    year = {2023},
    journal = {Journal on Technology and Persons with Disabilities},
    author = {Pitcher-Cooper, Charity and Seth, Manali and Kao, Benjamin and Coughlan, James M and Yoon, Ilmi},
    volume = {11},
    url = {https://youdescribe.org/}
}

@article{branje2012livedescribe,
  author    = {Carmen J. Branje and Deborah I. Fels},
  title     = {LiveDescribe: Can Amateur Describers Create High-Quality Audio Description?},
  journal   = {Journal of Visual Impairment \& Blindness},
  volume    = {106},
  number    = {3},
  pages     = {154--165},
  year      = {2012},
  doi       = {10.1177/0145482X1210600304}
}

@article{Wang2021TowardVideos,
    title = {{Toward automatic audio description generation for accessible videos}},
    year = {2021},
    journal = {Conference on Human Factors in Computing Systems - Proceedings},
    author = {Wang, Yujia and Liang, Wei},
    month = {5},
    publisher = {Association for Computing Machinery},
    url = {https://dl.acm.org/doi/10.1145/3411764.3445347},
    isbn = {9781450380966},
    doi = {10.1145/3411764.3445347/SUPPL{\_}FILE/3411764.3445347{\_}VIDEOPREVIEW.MP4}
}

@article{Rohrbach2017MovieDescription,
    title = {{Movie Description}},
    year = {2017},
    journal = {International Journal of Computer Vision},
    author = {Rohrbach, Anna and Torabi, Atousa and Rohrbach, Marcus and Tandon, Niket and Pal, Christopher and Larochelle, Hugo and Courville, Aaron and Schiele, Bernt},
    number = {1},
    month = {5},
    pages = {94--120},
    volume = {123},
    publisher = {Kluwer Academic PublishersPUB879Norwell, MA, USA},
    url = {https://dl.acm.org/doi/10.1007/s11263-016-0987-1},
    doi = {10.1007/S11263-016-0987-1},
    issn = {15731405},
    arxivId = {1605.03705}
}

@inproceedings{Yuksel2020Human-in-the-LoopUsers,
  title     = {Human-in-the-Loop Machine Learning to Increase Video Accessibility for Visually Impaired and Blind Users},
  author    = {Yuksel, Beste F. and Fazli, Pooyan and Mathur, Umang and Bisht, Vaishali and Kim, Soo Jung and Lee, Joshua Junhee and Jin, Seung Jung and Siu, Yue-Ting and Miele, Joshua A. and Yoon, Ilmi},
  booktitle = {Proceedings of the 2020 ACM Designing Interactive Systems Conference (DIS '20)},
  year      = {2020},
  address   = {Eindhoven, Netherlands},
  publisher = {Association for Computing Machinery},
  pages     = {47--60},
  doi       = {10.1145/3357236.3395433}
}

@article{Pavel2020Rescribe:Descriptions,
    title = {{Rescribe: Authoring and automatically editing audio descriptions}},
    year = {2020},
    journal = {UIST 2020 - Proceedings of the 33rd Annual ACM Symposium on User Interface Software and Technology},
    author = {Pavel, Amy and Reyes, Gabriel and Bigham, Jeffrey P.},
    month = {10},
    pages = {747--759},
    publisher = {Association for Computing Machinery, Inc},
    url = {https://dl.acm.org/doi/10.1145/3379337.3415864},
    isbn = {9781450375146},
    doi = {10.1145/3379337.3415864/SUPPL{\_}FILE/3379337.3415864.MP4}
}

@inproceedings{cheema2025describepro,
  author = {Cheema, Jawad and Ihorn, Stefan and Natalie, Rosiana and Hara, Kotaro and Pavel, Amy},
  title = {DescribePro: Collaborative Audio Description with Human-AI Interaction},
  booktitle = {The 27th International ACM SIGACCESS Conference on Computers and Accessibility (ASSETS '25)},
  year = {2025},
  location = {Denver, CO, USA},
  doi = {10.1145/3663547.3746320},
  publisher = {Association for Computing Machinery}
}

@inproceedings{liu2022crossa11y,
  author = {Liu, Xingyu "Bruce" and Wang, Ruolin and Li, Dingzeyu and Chen, Xiang 'Anthony' and Pavel, Amy},
  title = {CrossA11y: Identifying Video Accessibility Issues via Cross-Modal Grounding},
  booktitle = {Proceedings of the 35th Annual ACM Symposium on User Interface Software and Technology (UIST '22)},
  year = {2022},
  location = {Bend, OR, USA},
  doi = {10.1145/3526113.3545703},
  articleno = {43},
  numpages = {14},
  publisher = {Association for Computing Machinery}
}

@inproceedings{natalie2023supporting,
  author = {Natalie, Rosiana and Tseng, Joshua and Kacorri, Hernisa and Hara, Kotaro},
  title = {Supporting Novices Author Audio Descriptions via Automatic Feedback},
  booktitle = {Proceedings of the 2023 CHI Conference on Human Factors in Computing Systems (CHI '23)},
  year = {2023},
  location = {Hamburg, Germany},
  doi = {10.1145/3544548.3581023},
  articleno = {77},
  numpages = {18},
  publisher = {Association for Computing Machinery}
}

@article{chu2024llmad,
  author = {Chu, Yingqiang and Li, Yong and Yoon, Joon-Young and Park, Byung-Kwon},
  title = {LLM-AD: Large Language Model based Audio Description System},
  journal = {arXiv preprint arXiv:2405.00983},
  year = {2024}
}

@inproceedings{Ye2024MMAD:Description,
  title     = {{MMAD}: Multi-modal Movie Audio Description},
  author    = {Ye, Xiaojun and Chen, Junhao and Li, Xiang and Xin, Haidong and Li, Chao and Zhou, Sheng and Bu, Jiajun},
  booktitle = {Proceedings of the 2024 Joint International Conference on Computational Linguistics, Language Resources and Evaluation ({LREC-COLING} 2024)},
  year      = {2024},
  address   = {Torino, Italy},
  publisher = {ELRA and ICCL},
  pages     = {11415--11428},
  url       = {https://aclanthology.org/2024.lrec-main.998/}
}

@article{Li2025VideoA11y:Description,
    title = {{VideoA11y: Method and Dataset for Accessible Video Description}},
    year = {2025},
    journal = {Conference on Human Factors in Computing Systems - Proceedings },
    author = {Li, Chaoyu and Padmanabhuni, Sid and Cheema, Maryam S. and Seifi, Hasti and Fazli, Pooyan},
    month = {4},
    publisher = {Association for Computing Machinery},
    url = {/doi/pdf/10.1145/3706598.3714096?download=true},
    isbn = {9798400713941},
    doi = {10.1145/3706598.3714096/SUPPL{\_}FILE/PN2974-TALK-VIDEO.MP4}
}

@article{Cheema2024DescribeIndividuals,
    title = {{Describe Now: User-Driven Audio Description for Blind and Low Vision Individuals}},
    year = {2024},
    author = {Cheema, Maryam and Seifi, Hasti and Fazli, Pooyan},
    month = {11},
    url = {http://arxiv.org/abs/2411.11835},
    arxivId = {2411.11835v1}
}

@inproceedings{LaubliAssessingEnvironment,
  title     = {Assessing Post-Editing Efficiency in a Realistic Translation Environment},
  author    = {L{\"a}ubli, Samuel and Fishel, Mark and Massey, Gary and Ehrensberger-Dow, Maureen and Volk, Martin},
  editor    = {O'Brien, Sharon and Simard, Michel and Specia, Lucia},
  booktitle = {Proceedings of the {MT} {Summit} {XIV} Workshop on Post-editing Technology and Practice ({WPTP} 2013)},
  year      = {2013},
  month     = sep,
  address   = {Nice, France},
  pages     = {83--91},
  url       = {https://aclanthology.org/2013.mtsummit-wptp.10/}
}

@article{Green2013TheTranslation,
    title = {{The efficacy of human post-editing for language translation}},
    year = {2013},
    journal = {Conference on Human Factors in Computing Systems - Proceedings},
    author = {Green, Spence and Heer, Jeffrey and Manning, Christopher D.},
    pages = {439--448},
    url = {/doi/pdf/10.1145/2470654.2470718?download=true},
    isbn = {9781450318990},
    doi = {10.1145/2470654.2470718}
}

@misc{2024DescribedDCMP,
    title = {{Described and Captioned Media Program (DCMP)}},
    year = {2024},
    booktitle = {Description Key for Educational Media},
    url = {https://dcmp.org/learn/descriptionkey}
}

@misc{NationalCenterforAccessibleMedia2017AccessibleGuidelines,
    title = {{Accessible Digital Media Guidelines}},
    year = {2017},
    author = {{National Center for Accessible Media}},
    url = {https://ncam.wgbh.org}
}

@misc{YouDescribeYouDescribe.Https://www.youdescribe.org/,
    title = {{YouDescribe. https://www.youdescribe.org/}},
    author = {{YouDescribe}}
}

@article{VanDaele2024MakingSummaries,
    title = {{Making Short-Form Videos Accessible with Hierarchical Video Summaries}},
    year = {2024},
    journal = {Conference on Human Factors in Computing Systems - Proceedings},
    author = {Van Daele, Tess and Iyer, Akhil and Zhang, Yuning and Derry, Jalyn C. and Huh, Mina and Pavel, Amy},
    month = {2},
    pages = {17},
    volume = {1},
    publisher = {Association for Computing Machinery},
    url = {http://arxiv.org/abs/2402.10382 http://dx.doi.org/10.1145/3613904.3642839},
    doi = {10.1145/3613904.3642839},
    arxivId = {2402.10382v1}
}

@inproceedings{ihorn2023potential,
  author = {Stangl, Abigale and Ihorn, Shasta and Siu, Yue-Ting and Bodi, Aditya and Castanon, Mar and Narins, Lothar D. and Yoon, Ilmi},
  title = {The Potential of a Visual Dialogue Agent in a Tandem Automated Audio Description System for Videos},
  booktitle = {Proceedings of the 25th International ACM SIGACCESS Conference on Computers and Accessibility (ASSETS '23)},
  year = {2023},
  publisher = {Association for Computing Machinery},
  doi = {10.1145/3597638.3608402}
}

@article{Bodi2021AutomatedUsers,
    title = {{Automated Video Description for Blind and Low Vision Users}},
    year = {2021},
    journal = {Conference on Human Factors in Computing Systems - Proceedings},
    author = {Bodi, Aditya and Fazli, Pooyan and Ihorn, Shasta and Siu, Yue Ting and Scott, Andrew T. and Narins, Lothar and Kant, Yash and Das, Abhishek and Yoon, Ilmi},
    month = {5},
    publisher = {Association for Computing Machinery},
    url = {https://dl.acm.org/doi/10.1145/3411763.3451810},
    isbn = {9781450380959},
    doi = {10.1145/3411763.3451810}
}

@inproceedings{li2025vidhalluc,
  author = {Li, Chaoyu and Im, Eun Woo and Fazli, Pooyan},
  title = {VidHalluc: Evaluating Temporal Hallucinations in Multimodal Large Language Models for Video Understanding},
  booktitle = {Proceedings of the IEEE/CVF Conference on Computer Vision and Pattern Recognition (CVPR)},
  year = {2025},
  pages = {13723--13733}
}

@article{Chang2022OmniScribe:Videos,
    title = {{OmniScribe: Authoring Immersive Audio Descriptions for 360{${}^\circ$} Videos}},
    year = {2022},
    journal = {UIST 2022 - Proceedings of the 35th Annual ACM Symposium on User Interface Software and Technology},
    author = {Chang, Ruei Che and Ting, Chao Hsien and Hung, Chia Sheng and Lee, Wan Chen and Chen, Liang Jin and Chao, Yu Tzu and Chen, Bing Yu and Guo, Anhong},
    month = {10},
    publisher = {Association for Computing Machinery, Inc},
    url = {https://dl.acm.org/doi/10.1145/3526113.3545613},
    isbn = {9781450393201},
    doi = {10.1145/3526113.3545613}
}

@inproceedings{huh2023avscript,
  author = {Huh, Mina and Yang, Saelyne and Peng, Yi-Hao and Chen, Xiang 'Anthony' and Kim, Young-Ho and Pavel, Amy},
  title = {AVscript: Accessible Video Editing with Audio-Visual Scripts},
  booktitle = {Proceedings of the 2023 CHI Conference on Human Factors in Computing Systems (CHI '23)},
  year = {2023},
  articleno = {796},
  numpages = {17},
  publisher = {Association for Computing Machinery},
  address = {New York, NY, USA},
  doi = {10.1145/3544548.3581494}
}

@article{bergin2025automating,
  author = {Bergin, Daniel and Oppegaard, Brett},
  title = {Automating Media Accessibility: An Approach for Analyzing Audio Description Across Generative Artificial Intelligence Algorithms},
  journal = {Technical Communication Quarterly},
  volume = {34},
  number = {2},
  pages = {169--184},
  year = {2025},
  doi = {10.1080/10572252.2024.2372771},
  publisher = {Routledge}
}

@article{walczak2017creative,
  author = {Walczak, Agnieszka and Fryer, Louise},
  title = {Creative Description: The Impact of Audio Description Style on Presence in Visually Impaired Audiences},
  journal = {British Journal of Visual Impairment},
  volume = {35},
  number = {1},
  pages = {6--17},
  year = {2017},
  doi = {10.1177/0264619616661603}
}

@article{schaefferlacroix2023beyond,
  title   = {Beyond Objectivity in Audio Description: New Practices and Perspectives},
  author  = {Schaeffer-Lacroix, Eva and Reviers, Nina and Di Giovanni, Elena},
  journal = {Journal of Audiovisual Translation},
  volume  = {6},
  number  = {2},
  pages   = {1--7},
  year    = {2023},
  doi     = {10.47476/jat.v6i2.2023.309}
}

@article{xie2024autoad,
  author = {Xie, Junyu and Han, Tengda and Bain, Max and Nagrani, Arsha and Varol, G\"{u}l and Xie, Weidi and Zisserman, Andrew},
  title = {AutoAD-Zero: A Training-Free Framework for Zero-Shot Audio Description},
  journal = {arXiv preprint arXiv:2407.15850},
  year = {2024}
}

@article{lin2023mmvid,
  author = {Lin, Kevin and Ahmed, Faisal and Li, Linjie and Lin, Chung-Ching and Azarnasab, Ehsan and Yang, Zhengyuan and Wang, Jianfeng and Liang, Lin and Liu, Zicheng and Lu, Yumao and Liu, Ce and Li, Lijuan},
  title = {MM-VID: Advancing Video Understanding with GPT-4V(ision)},
  journal = {arXiv preprint arXiv:2310.19773},
  year = {2023}
}

@inproceedings{chen2024sharegpt4video,
  author = {Chen, Lin and Wei, Xilin and Li, Jinsong and Dong, Xiaoyi and Zhang, Pan and Zang, Yuhang and Chen, Zehui and Duan, Haodong and Lin, Bin and Tang, Zhenyu and Yuan, Li and Qiao, Yu and Lin, Dahua and Zhao, Feng and Wang, Jiaqi},
  title = {ShareGPT4Video: Improving Video Understanding and Generation with Better Captions},
  booktitle = {Advances in Neural Information Processing Systems 37 (NeurIPS 2024), Datasets and Benchmarks Track},
  year = {2024}
}

@article{daems2017identifying,
  author = {Daems, Joke and Vandepitte, Sonia and Hartsuiker, Robert J. and Macken, Lieve},
  title = {Identifying the Machine Translation Error Types with the Greatest Impact on Post-editing Effort},
  journal = {Frontiers in Psychology},
  volume = {8},
  pages = {1282},
  year = {2017},
  doi = {10.3389/fpsyg.2017.01282}
}

@inproceedings{dhillon2024shaping,
  author = {Dhillon, Paramveer S. and Molaei, Somayeh and Li, Jiaqi and Golub, Maximilian and Zheng, Shaochun and Robert, Lionel P.},
  title = {Shaping Human-AI Collaboration: Varied Scaffolding Levels in Co-writing with Language Models},
  booktitle = {Proceedings of the CHI Conference on Human Factors in Computing Systems (CHI '24)},
  year = {2024},
  publisher = {Association for Computing Machinery},
  doi = {10.1145/3613904.3642134}
}

@article{Schuhmann2022LAION-5B:Models,
    title = {{LAION-5B: An open large-scale dataset for training next generation image-text models}},
    year = {2022},
    author = {Schuhmann, Christoph and Beaumont, Romain and Vencu, Richard and Gordon, Cade and Wightman, Ross and Cherti, Mehdi and Coombes, Theo and Katta, Aarush and Mullis, Clayton and Wortsman, Mitchell and Schramowski, Patrick and Kundurthy, Srivatsa and Crowson, Katherine and Schmidt, Ludwig and Kaczmarczyk, Robert and Jitsev, Jenia},
    month = {10},
    url = {http://arxiv.org/abs/2210.08402},
    arxivId = {2210.08402}
}

@misc{googleTTS,
  author = {{Google Cloud}},
  title = {Cloud Text-to-Speech API},
  year = {2024},
  url = {https://cloud.google.com/text-to-speech},
  note = {Accessed: 2025}
}

@article{Fernandez-Torne2015TheCatalan,
  title   = {Text-to-Speech vs. Human Voiced Audio Descriptions: A Reception Study in Films Dubbed into {Catalan}},
  author  = {Fern{\'a}ndez-Torn{\'e}, Anna and Matamala, Anna},
  journal = {The Journal of Specialised Translation},
  volume  = {24},
  pages   = {61--88},
  year    = {2015},
  doi     = {10.26034/cm.jostrans.2015.323},
  url     = {https://www.jostrans.org/article/view/7708}
}

@article{Walczak2018VocalPresence,
    title = {{Vocal delivery of audio description by genre: measuring users’ presence}},
    year = {2018},
    journal = {Perspectives: Studies in Translatology},
    author = {Walczak, Agnieszka and Fryer, Louise},
    number = {1},
    month = {1},
    pages = {69--83},
    volume = {26},
    publisher = {Routledge},
    doi = {10.1080/0907676X.2017.1298634},
    issn = {17476623}
}

@inproceedings{chen2024bge,
  author = {Chen, Jianlv and Xiao, Shitao and Zhang, Peitian and Luo, Kun and Lian, Defu and Liu, Zheng},
  title = {M3-Embedding: Multi-Linguality, Multi-Functionality, Multi-Granularity Text Embeddings Through Self-Knowledge Distillation},
  booktitle = {Findings of the Association for Computational Linguistics: ACL 2024},
  pages = {2318--2335},
  year = {2024},
  address = {Bangkok, Thailand},
  publisher = {Association for Computational Linguistics}
}

@inproceedings{rivera2021luar,
  author = {Rivera-Soto, Rafael A. and Miano, Olivia Elizabeth and Ordonez, Juanita and Chen, Barry Y. and Khan, Aleem and Bishop, Marcus and Andrews, Nicholas},
  title = {Learning Universal Authorship Representations},
  booktitle = {Proceedings of the 2021 Conference on Empirical Methods in Natural Language Processing (EMNLP)},
  pages = {913--919},
  year = {2021},
  address = {Online and Punta Cana, Dominican Republic},
  publisher = {Association for Computational Linguistics}
}

@article{levenshtein1966binary,
  author = {Levenshtein, Vladimir I.},
  title = {Binary Codes Capable of Correcting Deletions, Insertions, and Reversals},
  journal = {Soviet Physics Doklady},
  volume = {10},
  number = {8},
  pages = {707--710},
  year = {1966}
}

@article{bradley2007qualitative,
  author = {Bradley, Elizabeth H. and Curry, Leslie A. and Devers, Kelly J.},
  title = {Qualitative Data Analysis for Health Services Research: Developing Taxonomy, Themes, and Theory},
  journal = {Health Services Research},
  volume = {42},
  number = {4},
  pages = {1758--1772},
  year = {2007},
  doi = {10.1111/j.1475-6773.2006.00684.x}
}

@inproceedings{kadoma2024ownership,
  author = {Kadoma, Kowe and Aubin Le Quere, Marianne and Fu, Xiyu Jenny and Munsch, Christin and Metaxa, Dana\"{e} and Naaman, Mor},
  title = {The Role of Inclusion, Control, and Ownership in Workplace AI-Mediated Communication},
  booktitle = {Proceedings of the CHI Conference on Human Factors in Computing Systems (CHI '24)},
  year = {2024},
  publisher = {Association for Computing Machinery},
  doi = {10.1145/3613904.3642650}
}

@misc{openai2024gpt4o,
  author = {{OpenAI}},
  title = {GPT-4o},
  year = {2024},
  url = {https://openai.com/index/hello-gpt-4o/},
  note = {Accessed: 2025}
}

@inproceedings{jakesch2023cowriting,
  author = {Jakesch, Maurice and Bhat, Advait and Buschek, Daniel and Zalmanson, Lior and Naaman, Mor},
  title = {Co-Writing with Opinionated Language Models Affects Users' Views},
  booktitle = {Proceedings of the 2023 CHI Conference on Human Factors in Computing Systems (CHI '23)},
  year = {2023},
  publisher = {Association for Computing Machinery},
  doi = {10.1145/3544548.3581196}
}

@inproceedings{agarwal2025homogenize,
  author = {Agarwal, Dhruv and Naaman, Mor and Vashistha, Aditya},
  title = {AI Suggestions Homogenize Writing Toward Western Styles and Diminish Cultural Nuances},
  booktitle = {Proceedings of the 2025 CHI Conference on Human Factors in Computing Systems (CHI '25)},
  year = {2025},
  location = {Yokohama, Japan},
  publisher = {Association for Computing Machinery},
  doi = {10.1145/3706598.3713564}
}

@article{sakoe1978dtw,
  author = {Sakoe, Hiroaki and Chiba, Seibi},
  title = {Dynamic Programming Algorithm Optimization for Spoken Word Recognition},
  journal = {IEEE Transactions on Acoustics, Speech, and Signal Processing},
  volume = {26},
  number = {1},
  pages = {43--49},
  year = {1978},
  doi = {10.1109/TASSP.1978.1163055}
}

@incollection{hart1988nasa,
  author = {Hart, Sandra G. and Staveland, Lowell E.},
  title = {Development of {NASA-TLX} (Task Load Index): Results of Empirical and Theoretical Research},
  booktitle = {Human Mental Workload},
  editor = {Hancock, Peter A. and Meshkati, Najmedin},
  publisher = {North-Holland},
  address = {Amsterdam},
  pages = {139--183},
  year = {1988},
  doi = {10.1016/S0166-4115(08)62386-9}
}

@inproceedings{lee2024designspace,
  author = {Lee, Mina and Gero, Katy Ilonka and Chung, John Joon Young and Buckingham Shum, Simon and Raheja, Vipul and Shen, Hua and Venugopalan, Subhashini and Wambsganss, Thiemo and Zhou, David and Alghamdi, Emad A. and others},
  title = {A Design Space for Intelligent and Interactive Writing Assistants},
  booktitle = {Proceedings of the CHI Conference on Human Factors in Computing Systems (CHI '24)},
  year = {2024},
  publisher = {Association for Computing Machinery},
  address = {New York, NY, USA},
  doi = {10.1145/3613904.3642697}
}

@article{TomarSuramya2006ConvertingFFmpeg,
    title = {{Converting video formats with FFmpeg}},
    year = {2006},
    journal = {Linux Journal},
    author = {{TomarSuramya}},
    month = {6},
    publisher = {Belltown MediaPUB6702Houston, TX},
    url = {https://dl.acm.org/doi/10.5555/1134782.1134792},
    doi = {10.5555/1134782.1134792}
}

@article{jamison2003turnover,
  author  = {Jamison, Irma Browne},
  title   = {Turnover and Retention among Volunteers in Human Service Agencies},
  journal = {Review of Public Personnel Administration},
  volume  = {23},
  number  = {2},
  pages   = {114--132},
  year    = {2003},
  doi     = {10.1177/0734371X03023002003}
}

@book{krings2001repairing,
  author    = {Krings, Hans P.},
  editor    = {Koby, Geoffrey S.},
  title     = {Repairing Texts: Empirical Investigations of Machine
               Translation Post-Editing Processes},
  publisher = {Kent State University Press},
  address   = {Kent, OH},
  year      = {2001}
}

@inproceedings{snover2006ter,
  author    = {Snover, Matthew and Dorr, Bonnie and Schwartz, Richard and
               Micciulla, Linnea and Makhoul, John},
  title     = {A Study of Translation Edit Rate with Targeted Human
               Annotation},
  booktitle = {Proceedings of the 7th Conference of the Association for
               Machine Translation in the Americas},
  pages     = {223--231},
  year      = {2006}
}

@inproceedings{koponen2012comparing,
  author    = {Koponen, Maarit},
  title     = {Comparing Human Perceptions of Post-Editing Effort with
               Post-Editing Operations},
  booktitle = {Proceedings of the Seventh Workshop on Statistical Machine
               Translation},
  pages     = {181--190},
  address   = {Montr\'eal, Canada},
  publisher = {Association for Computational Linguistics},
  year      = {2012}
}

@book{fryer2016introduction,
  author    = {Fryer, Louise},
  title     = {An Introduction to Audio Description: A Practical Guide},
  publisher = {Routledge},
  address   = {London},
  year      = {2016}
}

@book{snyder2014visual,
  author    = {Snyder, Joel},
  title     = {The Visual Made Verbal: A Comprehensive Training Manual and
               Guide to the History and Applications of Audio Description},
  publisher = {American Council of the Blind},
  address   = {Arlington, VA},
  year      = {2014}
}

@article{draxler2024ghostwriter,
  author  = {Draxler, Fiona and Werner, Anna and Lehmann, Florian and
             Hoppe, Matthias and Schmidt, Albrecht and Buschek, Daniel and
             Welsch, Robin},
  title   = {The {AI} Ghostwriter Effect: When Users Do Not Perceive
             Ownership of {AI}-Generated Text but Self-Declare as Authors},
  journal = {ACM Transactions on Computer-Human Interaction},
  volume  = {31},
  number  = {2},
  articleno = {25},
  year    = {2024},
  doi     = {10.1145/3637875}
}

@inproceedings{lee2022coauthor,
  author    = {Lee, Mina and Liang, Percy and Yang, Qian},
  title     = {{CoAuthor}: Designing a Human-{AI} Collaborative Writing
               Dataset for Exploring Language Model Capabilities},
  booktitle = {Proceedings of the 2022 CHI Conference on Human Factors in
               Computing Systems},
  articleno = {388},
  numpages  = {19},
  year      = {2022},
  doi       = {10.1145/3491102.3502030}
}

@inproceedings{natalie2021viscene,
  author    = {Natalie, Rosiana and Loh, Jolene and Tan, Huei Suen and
               Tseng, Joshua and Chan, Ian Luke Yi-Ren and Jarjue, Ebrima H.
               and Kacorri, Hernisa and Hara, Kotaro},
  title     = {The Efficacy of Collaborative Authoring of Video Scene
               Descriptions},
  booktitle = {Proceedings of the 23rd International ACM SIGACCESS
               Conference on Computers and Accessibility (ASSETS '21)},
  year      = {2021},
  doi       = {10.1145/3441852.3471201}
}

@misc{lintoai2023whispertimestamped,
  title={whisper-timestamped},
  author={Louradour, J{\'e}r{\^o}me},
  journal={GitHub repository},
  year={2023},
  publisher={GitHub},
  howpublished = {\url{https://github.com/linto-ai/whisper-timestamped}}
}

@article{JSSv031i07,
  title={Computing and Visualizing Dynamic Time Warping Alignments in R: The dtw Package},
  author={Giorgino, Toni},
  journal={Journal of Statistical Software},
  year={2009},
  volume={31},
  number={7},
  doi={10.18637/jss.v031.i07}
}

@article{Radford2022RobustSupervision,
  title     = {Robust Speech Recognition via Large-Scale Weak Supervision},
  year      = {2022},
  journal   = {Proceedings of Machine Learning Research},
  author    = {Radford, Alec and Kim, Jong Wook and Xu, Tao and Brockman, Greg and McLeavey, Christine and Sutskever, Ilya},
  month     = {12},
  pages     = {28492--28518},
  volume    = {202},
  url       = {https://arxiv.org/pdf/2212.04356},
  eprint    = {2212.04356}
}

@inproceedings{Radford2021CLIP,
  author    = {Radford, Alec and Kim, Jong Wook and Hallacy, Chris and Ramesh, Aditya and Goh, Gabriel and Agarwal, Sandhini and Sastry, Girish and Askell, Amanda and Mishkin, Pamela and Clark, Jack and Krueger, Gretchen and Sutskever, Ilya},
  title     = {Learning Transferable Visual Models From Natural Language Supervision},
  booktitle = {Proceedings of the 38th International Conference on Machine Learning},
  series    = {Proceedings of Machine Learning Research},
  volume    = {139},
  pages     = {8748--8763},
  year      = {2021},
  publisher = {PMLR}
}

@inproceedings{arnold2020predictive,
  author    = {Arnold, Kenneth C. and Chaunce, Krysta and Gajos, Krzysztof Z.},
  title     = {Predictive Text Encourages Predictable Writing},
  booktitle = {Proceedings of the 25th International Conference on Intelligent User Interfaces},
  series    = {IUI '20},
  year      = {2020},
  publisher = {Association for Computing Machinery},
  address   = {New York, NY, USA},
  doi = {10.1145/3377325.3377523}
}

@inproceedings{bhat2023nextphrase,
  author    = {Bhat, Advait and Agashe, Saaket and Oberoi, Parth and Mohile, Niharika and Jangir, Ravi and Joshi, Anirudha},
  title     = {Interacting with Next-Phrase Suggestions: How Suggestion Systems Aid and Influence the Cognitive Processes of Writing},
  booktitle = {Proceedings of the 28th International Conference on Intelligent User Interfaces},
  series    = {IUI '23},
  year      = {2023},
  pages     = {436--452},
  publisher = {Association for Computing Machinery},
  address   = {New York, NY, USA},
  doi       = {10.1145/3581641.3584060}
}

@inproceedings{schemmer2023reliance,
  author    = {Schemmer, Max and Kuehl, Niklas and Benz, Carina and Bartos, Andrea and Satzger, Gerhard},
  title     = {Appropriate Reliance on {AI} Advice: Conceptualization and the Effect of Explanations},
  booktitle = {Proceedings of the 28th International Conference on Intelligent User Interfaces},
  series    = {IUI '23},
  year      = {2023},
  pages     = {410--422},
  publisher = {Association for Computing Machinery},
  address   = {New York, NY, USA},
  doi       = {10.1145/3581641.3584066}
}

@inproceedings{hazra2026accepted,
  author    = {Hazra, Sanchaita and Lee, Doeun and Majumder, Bodhisattwa Prasad and Kumar, Sachin},
  title     = {Accepted with Minor Revisions: Value of {AI}-Assisted Scientific Writing},
  booktitle = {Proceedings of the 31st International Conference on Intelligent User Interfaces},
  series    = {IUI '26},
  year      = {2026},
  pages     = {1--22},
  publisher = {Association for Computing Machinery},
  address   = {New York, NY, USA},
  doi       = {10.1145/3742413.3789140}
}

@inproceedings{sun2024reviewflow,
  author    = {Sun, Lu and Chan, Aaron and Chang, Yun Seo and Dow, Steven P.},
  title     = {ReviewFlow: Intelligent Scaffolding to Support Academic Peer Reviewing},
  booktitle = {Proceedings of the 29th International Conference on Intelligent User Interfaces},
  series    = {IUI '24},
  year      = {2024},
  pages     = {120--137},
  publisher = {Association for Computing Machinery},
  address   = {New York, NY, USA},
  doi       = {10.1145/3640543.3645159}
}

@inproceedings{neshaei2024peer,
  author    = {Neshaei, Seyed Parsa and Rietsche, Roman and Su, Xiaotian and Wambsganss, Thiemo},
  title     = {Enhancing Peer Review with {AI}-Powered Suggestion Generation Assistance: Investigating the Design Dynamics},
  booktitle = {Proceedings of the 29th International Conference on Intelligent User Interfaces},
  series    = {IUI '24},
  year      = {2024},
  pages     = {88--102},
  publisher = {Association for Computing Machinery},
  address   = {New York, NY, USA},
  doi       = {10.1145/3640543.3645169}
}

@inproceedings{clark2018creative,
  author    = {Clark, Elizabeth and Ross, Anne Spencer and Tan, Chenhao and Ji, Yangfeng and Smith, Noah A.},
  title     = {Creative Writing with a Machine in the Loop: Case Studies on Slogans and Stories},
  booktitle = {Proceedings of the 23rd International Conference on Intelligent User Interfaces},
  series    = {IUI '18},
  year      = {2018},
  pages     = {329--340},
  publisher = {Association for Computing Machinery},
  address   = {New York, NY, USA},
  doi       = {10.1145/3172944.3172983}
}

@inproceedings{yuan2022wordcraft,
  author    = {Yuan, Ann and Coenen, Andy and Reif, Emily and Ippolito, Daphne},
  title     = {Wordcraft: Story Writing With Large Language Models},
  booktitle = {Proceedings of the 27th International Conference on Intelligent User Interfaces},
  series    = {IUI '22},
  year      = {2022},
  pages     = {841--852},
  publisher = {Association for Computing Machinery},
  address   = {New York, NY, USA},
  doi       = {10.1145/3490099.3511105}
}

@inproceedings{seo2026whose,
  author    = {Seo, Jwawon and Deldari, Elmira and Mentis, Helena M.},
  title     = {Whose Code Is It? How {AI} Autonomy Reshapes Ownership, Responsibility, and Disclosure in {AI}-Assisted Programming},
  booktitle = {Proceedings of the 31st International Conference on Intelligent User Interfaces},
  series    = {IUI '26},
  year      = {2026},
  pages     = {393--425},
  publisher = {Association for Computing Machinery},
  address   = {New York, NY, USA},
  doi       = {10.1145/3742413.3789121}
}

@inproceedings{nakano2026disclosure,
  author    = {Nakano, Hiroki and Takezawa, Jo and Matulic, Fabrice and Yang, Chi-Lan and Yatani, Koji},
  title     = {Understanding Reader Perception Shifts upon Disclosure of {AI} Authorship},
  booktitle = {Proceedings of the 31st International Conference on Intelligent User Interfaces},
  series    = {IUI '26},
  year      = {2026},
  pages     = {2131--2146},
  publisher = {Association for Computing Machinery},
  address   = {New York, NY, USA},
  doi       = {10.1145/3742413.3789076}
}

@article{rose1980rigid,
  author    = {Rose, Mike},
  title     = {Rigid Rules, Inflexible Plans, and the Stifling of Language: A Cognitivist Analysis of Writer's Block},
  journal   = {College Composition and Communication},
  volume    = {31},
  number    = {4},
  pages     = {389--401},
  year      = {1980},
  publisher = {National Council of Teachers of English}
}

@incollection{salway2007,
  author    = {Salway, Andrew},
  title     = {A Corpus-based Analysis of Audio Description},
  booktitle = {Media for All: Subtitling for the Deaf, Audio Description, and Sign Language},
  editor    = {D{\'i}az Cintas, Jorge and Orero, Pilar and Remael, Aline},
  publisher = {Rodopi},
  address   = {Amsterdam},
  pages     = {151--174},
  year      = {2007}
}

@article{piety2004,
  author  = {Piety, Philip J.},
  title   = {The Language System of Audio Description: An Investigation as a Discursive Process},
  journal = {Journal of Visual Impairment \& Blindness},
  volume  = {98},
  number  = {8},
  pages   = {453--469},
  year    = {2004}
}

@techreport{itc2000,
  author      = {{Independent Television Commission}},
  title       = {{ITC} Guidance on Standards for Audio Description},
  institution = {Independent Television Commission},
  address     = {London, UK},
  year        = {2000}
}

@misc{netflixad,
  author       = {{Netflix}},
  title        = {Audio Description Style Guide},
  howpublished = {\url{https://partnerhelp.netflixstudios.com/hc/en-us/articles/215510667}},
  note         = {Accessed August 2026}
}

@misc{wcag21,
  author       = {{World Wide Web Consortium}},
  title        = {Web Content Accessibility Guidelines ({WCAG}) 2.1, Success Criterion 1.2.7: Extended Audio Description},
  year         = {2018},
  howpublished = {\url{https://www.w3.org/TR/WCAG21/}}
}

@book{remael2015adlab,
  editor    = {Remael, Aline and Reviers, Nina and Vercauteren, Gert},
  title     = {Pictures Painted in Words: {ADLAB} Audio Description Guidelines},
  publisher = {EUT Edizioni Universit{\`a} di Trieste},
  address   = {Trieste, Italy},
  year      = {2015}
}

@inproceedings{hemmer2023delegation,
  author    = {Hemmer, Patrick and Westphal, Monika and Schemmer, Max and Vetter, Sebastian and V{\"o}ssing, Michael and Satzger, Gerhard},
  title     = {Human-AI Collaboration: The Effect of AI Delegation on Human Task Performance and Task Satisfaction},
  booktitle = {Proceedings of the 28th International Conference on Intelligent User Interfaces},
  series    = {IUI '23},
  year      = {2023},
  pages     = {453--463},
  publisher = {Association for Computing Machinery},
  address   = {New York, NY, USA},
  doi       = {10.1145/3581641.3584052}
}

@inproceedings{do2025youdescribe,
  author    = {Do, Lana and Mirani, Sanjay and Pitcher-Cooper, Charity and Nguyen, Xuan Duy Anh and Bairaboina, Alekya and Yoon, Ilmi},
  title     = {YouDescribe: Bridging AI Efficiency and Human Insight for Scalable Audio Description},
  booktitle = {Adjunct Proceedings of the 4th Annual Symposium on Human-Computer Interaction for Work},
  series    = {CHIWORK '25 Adjunct},
  year      = {2025},
  articleno = {10},
  pages     = {1--7},
  publisher = {Association for Computing Machinery},
  address   = {New York, NY, USA},
  doi       = {10.1145/3707640.3731916}
}
\clearpage
\appendix

\section{GenAD Prompt Templates}
\label{appendix:prompts}
 
This section contains the full prompt templates used in our system, including the system prompt with audio description guidelines, scene-level generation, optimization, and extended description filtering prompts.

\subsection{System Prompt}
 
The system prompt establishes the audio describer role and injects domain-specific guidelines that govern all downstream generation steps.
 
\begin{figure}[H]
\centering
\begin{lstlisting}[style=prompt, title={\footnotesize\textbf{Guidelines String}}]
AUDIO DESCRIPTION GUIDELINES:
- Describe what you see in a concise, factual manner.
- Always read on-screen text exactly as it appears.
- Be factual, objective, and precise in your
  descriptions.
- Use proper terminology and names from the context
  when possible.
- Match the tone and mood of the video.
- Do not over-describe -- less is more.
- Do not interpret or editorialize about what you see.
- Do not give away surprises before they happen.
 
CHARACTER IDENTIFICATION:
- When you recognize a character from the context,
  ALWAYS use their specific name.
- Before each scene, carefully review context to
  identify all named characters.
- Use the most specific identification possible based
  on the context information.
\end{lstlisting}
 
\vspace{4pt}
 
\begin{lstlisting}[style=prompt, title={\footnotesize\textbf{System Message}}]
You are an expert audio describer creating descriptions for videos to enhance accessibility for blind and low-vision users. Your task is to generate video descriptions that are true,
contextually meaningful, useful, well-timed, and
polished. You must follow the audio description guidelines below.

{guidelines}

\end{lstlisting}
\caption{System prompt: the \texttt{guidelines} string is injected into the system message via the \texttt{\{guidelines\}} placeholder.}
\label{fig:system-prompt}
\end{figure}

\subsection{Scene-Level Generation Prompt}
 
This prompt is issued once per scene. It receives the scene duration and narrative context, then requests structured JSON output for both on-screen text events and visual description events.
 
\begin{figure}[H]
\centering
\begin{lstlisting}[style=prompt]
SCENE DURATION: {scene_duration:.2f} seconds
 
CONTEXT:
{context}
 
You are analyzing a video scene. Identify specific
characters, locations, and any important elements
mentioned in the context.
 
First, generate a JSON array of Text on Screen events
("type": "Text on Screen"):
- Capture ALL visible on-screen text.
- DO NOT include transcript or dialogue.
- For each text event, include the EXACT start_time
  in seconds when the text appears.
 
INCLUDE: Titles, headings, names; informational text;
important dates or events.
EXCLUDE: Brand logos and watermarks; network logos;
social media handles; copyright notices.
 
Second, generate a JSON array of Visual events:
- Provide contextually rich visual description of the
  scene using concise wording.
- Describe each action in detail.
- ALWAYS use specific character names from context
  (not "person" or "woman").
- Focus on key actions, settings, and objects not
  mentioned in a previous description.
- Include clear start times for each visual event.
- DO NOT describe Text on Screen.
 
OUTPUT FORMAT (JSON array):
  - start_time  (in seconds)
  - type        ("Text on Screen" or "Visual")
  - text        (description or on-screen text)
\end{lstlisting}
\caption{Prompt for generating scene-level descriptions and text events.}
\label{fig:scene-gen-prompt}
\end{figure}

\subsection{Inline Optimization Prompt}
 
When the estimated speech duration of a generated description exceeds the available silence gap, the following prompt asks the model to condense the text to fit the timing constraint.
 
\begin{figure}[H]
\centering
\begin{lstlisting}[style=prompt]
You are optimizing a set of visual descriptions for
a video.
 
ORIGINAL DESCRIPTIONS: "{combined_text}"
AVAILABLE TIME: {available_duration:.2f} seconds
 
TASK:
Combine and condense these descriptions to fit within
{available_duration:.2f} seconds of spoken audio.
 
GUIDELINES:
- Create a coherent, flowing description.
- Maintain the action order and use concise but
  natural language.
- Ensure the final description can be spoken within
  the time limit.
 
OUTPUT: Provide only the optimized description text,
without explanations.
\end{lstlisting}
\caption{Prompt for inline description optimization.}
\label{fig:inline-opt-prompt}
\end{figure}

\subsection{Retry Optimization Prompt}
 
If the first optimization attempt still exceeds the time budget (verified via TTS), this follow-up prompt provides the measured overshoot and requests further reduction.
 
\begin{figure}[H]
\centering
\begin{lstlisting}[style=prompt]
You are optimizing visual descriptions for a video.
 
PREVIOUS ATTEMPT: "{optimized_text}"
 
This description takes {tts_duration:.2f} seconds to
speak, but only {available_duration:.2f} seconds are
available. Reduce by {tts_duration -
available_duration:.2f} seconds.
 
TASK:
Create a SHORTER version that fits within the time
limit.
 
GUIDELINES:
- Keep the most critical visual elements.
- Eliminate redundant details.
- Use concise but natural language.
 
OUTPUT: Provide only the shortened description,
nothing else.
\end{lstlisting}
\caption{Prompt for retrying a failed inline optimization.}
\label{fig:retry-opt-prompt}
\end{figure}

\subsection{Extended Description Filtering Prompt}
 
For scenes where inline timing is insufficient, this prompt evaluates candidate descriptions against the transcript and cumulative context to decide whether an extended (pause-the-video) description is warranted.
 
\begin{figure}[H]
\centering
\begin{lstlisting}[style=prompt]
You are an accessibility expert selecting ONE visual
description per scene to convert to audio description
for blind and low-vision users.
 
CONTEXT:
- AD should be minimal and only interrupt audio when
  necessary.
- Spoken transcript is the primary information source.
- Use AD only when critical visual information is
  missing from the audio.
 
INPUT:
CURRENT SCENE TRANSCRIPT: {transcript_text}
CUMULATIVE TRANSCRIPT:    {cumulative_transcript}
CUMULATIVE DESCRIPTION:   {previous_desc_text}
VISUAL DESCRIPTIONS TO EVALUATE: {clip['text']}
 
EVALUATION CRITERIA (mark necessary = true if any):
- Conveys visual details absent from the audio.
- Describes important silent actions or key events.
- Identifies characters or locations that are
  otherwise ambiguous.
- Introduces a novel or distinct visual element.
- Notes scene changes or unannounced time jumps.
 
OUTPUT FORMAT (JSON array, one item per description):
  - id         (index of the description)
  - necessary  (true or false)
  - reason     (short explanation)
\end{lstlisting}
\caption{Prompt for filtering extended audio description candidates.}
\label{fig:ext-filter-prompt}
\end{figure}
\section{AI-Generated AD Preview Interface}
\label{appendix:ai-review}

\begin{figure*}[t]
  \centering
  \includegraphics[width=0.75\textwidth]{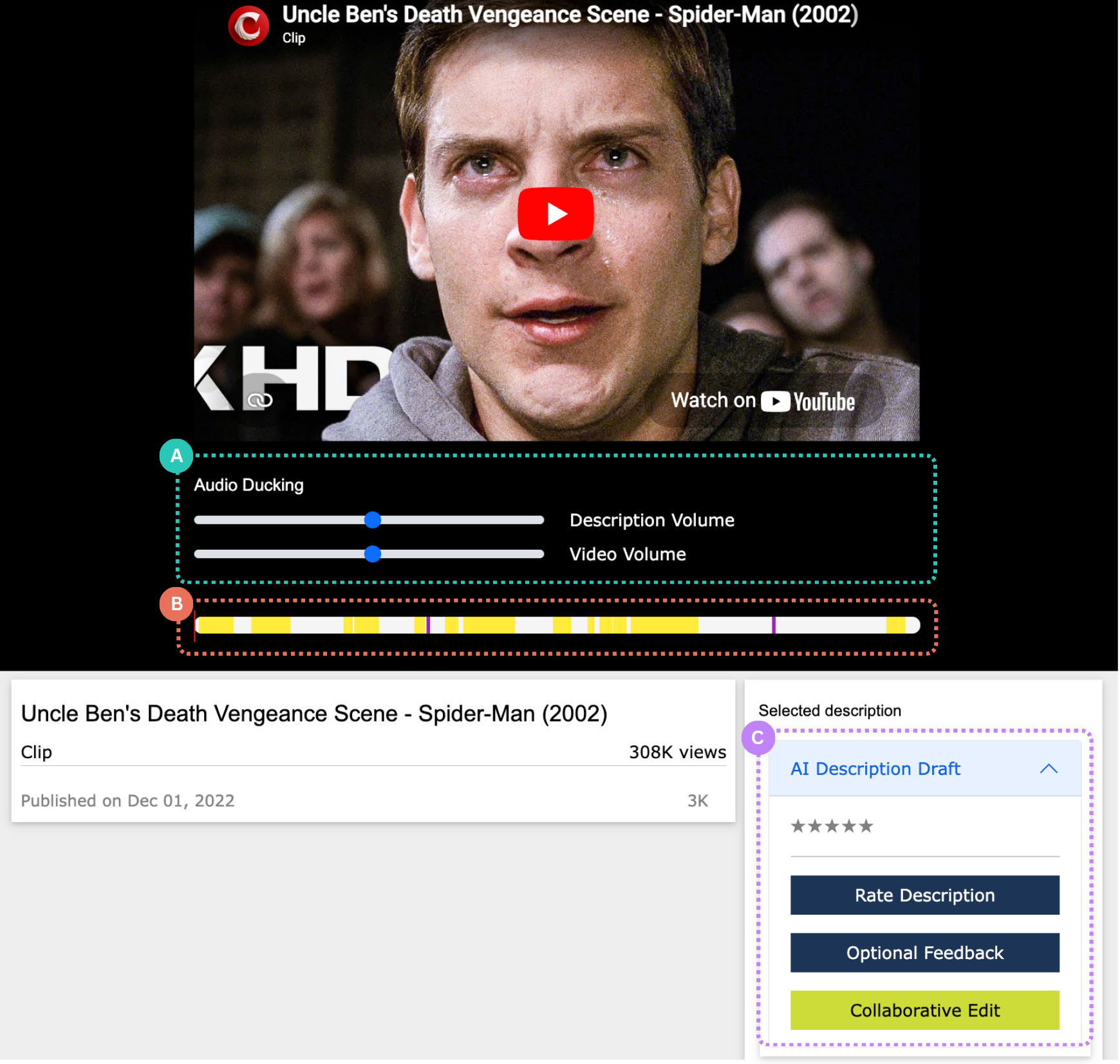}
  \caption{Video preview with AI-generated AD. Users watch the video with audio ducking controls~(A) and view the timeline of inserted clips and dialogue~(B; yellow~=~inline, purple~=~extended). From the selected description panel~(C), users can rate the AI-generated description, provide optional feedback, or begin collaborative editing in RefineAD.}
  \label{fig:preview-ai}
\end{figure*}

Figure~\ref{fig:preview-ai} shows the preview interface for AI-generated drafts. Once GenAD completes a draft, any platform user can watch the video with synchronized descriptions, rate individual clips, and open the draft in RefineAD for collaborative editing. Because the draft is stored with the video, repeat requests for the same video are avoided.

\FloatBarrier

\section{From-Scratch Authoring Interface}
\label{appendix:from-scratch}

\begin{figure*}[t]
  \centering
  \begin{subfigure}[t]{0.6\textwidth}
    \centering
    \includegraphics[width=\linewidth]{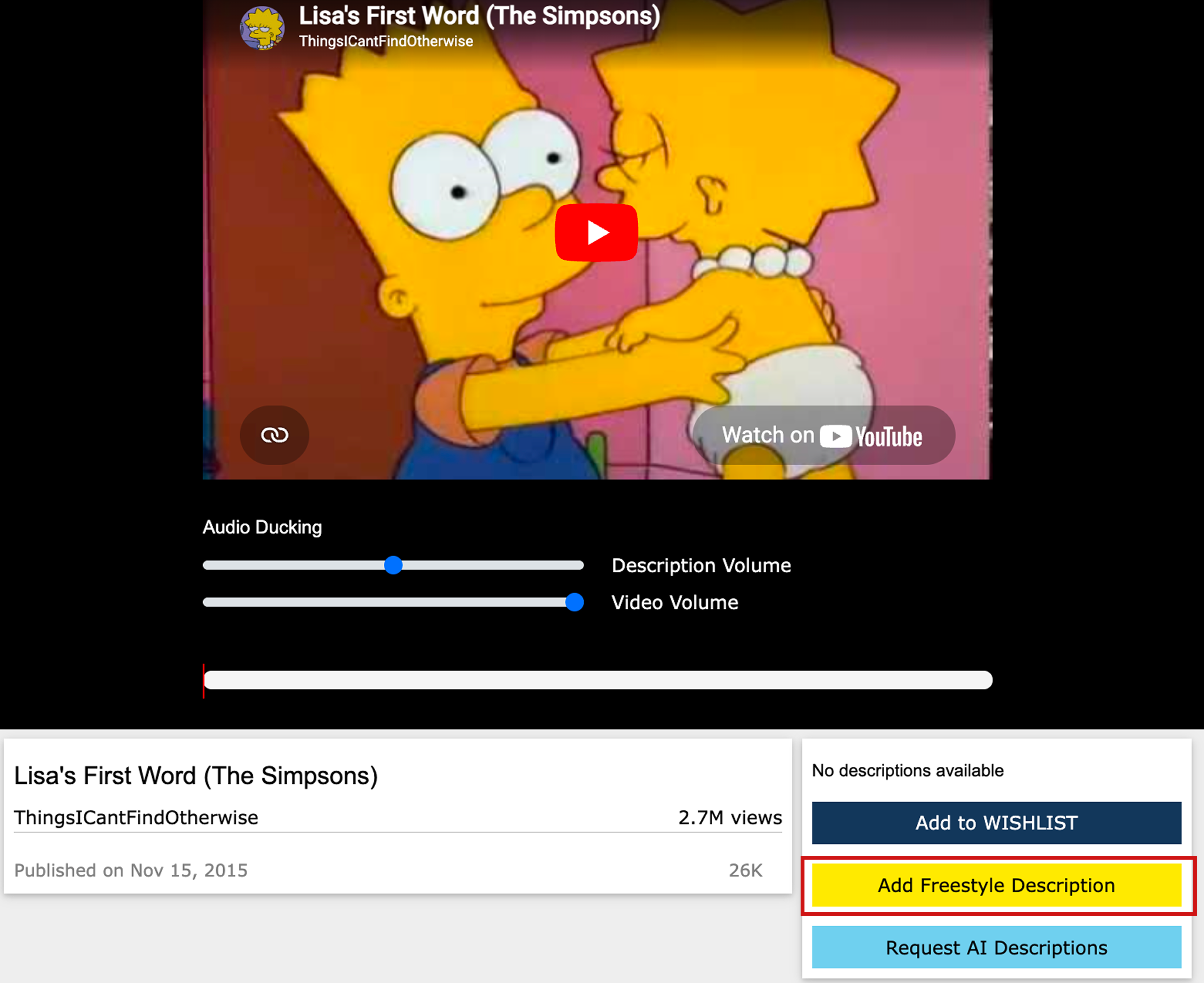}
    \caption{Video landing page for an undescribed video, with ``Add Freestyle Description'' highlighted.}
    \label{fig:from-scratch-landing}
  \end{subfigure}

  \vspace{0.75em}

  \begin{subfigure}[t]{0.6\textwidth}
    \centering
    \includegraphics[width=\linewidth]{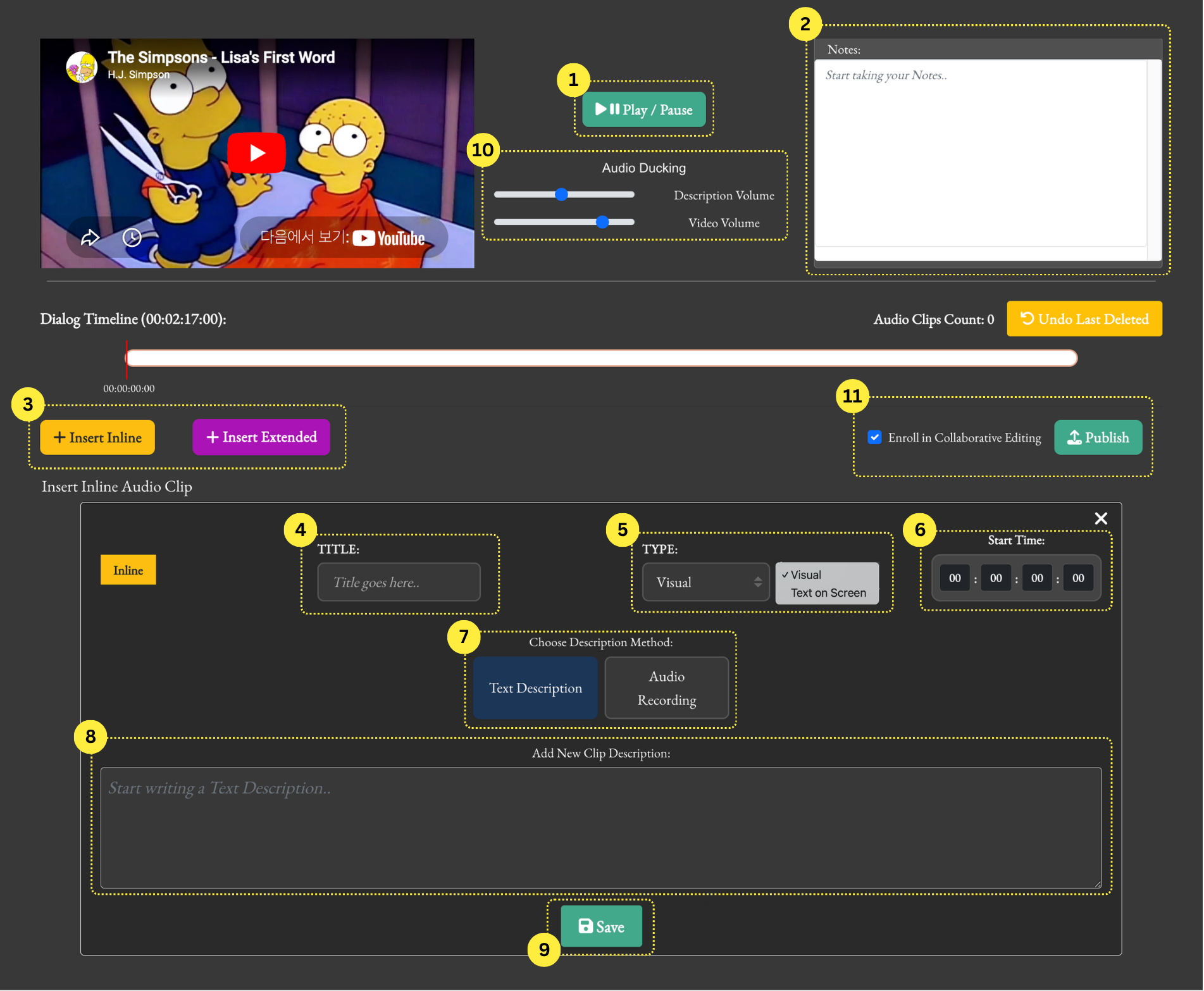}
    \caption{From-scratch authoring interface. Describers preview the video with play/pause controls~(1) and take notes~(2). They insert inline or extended descriptions~(3; yellow~=~inline, purple~=~extended), add a title~(4), select the description type~(5), adjust the start time~(6), choose a description method~(7), write the description text~(8), and save~(9). Audio ducking is controlled via~(10); the video is published, optionally with collaborative editing enabled, via~(11).}
    \label{fig:from-scratch-editor}
  \end{subfigure}
  \caption{From-scratch authoring workflow.}
  \label{fig:from-scratch}
\end{figure*}

Users can search for any YouTube video on the platform. When no descriptions exist, the landing page offers two actions: add a freestyle description or request AI-generated descriptions. Selecting ``Add Freestyle Description'' (Figure~\ref{fig:from-scratch-landing}) opens the from-scratch authoring interface (Figure~\ref{fig:from-scratch-editor}), where describers compose and publish descriptions without AI assistance.

\FloatBarrier
\section{Study Interface and Condition Assignment}
\label{appendix:study-design}

\begin{figure}[!htb]
  \centering
  \includegraphics[width=\columnwidth]{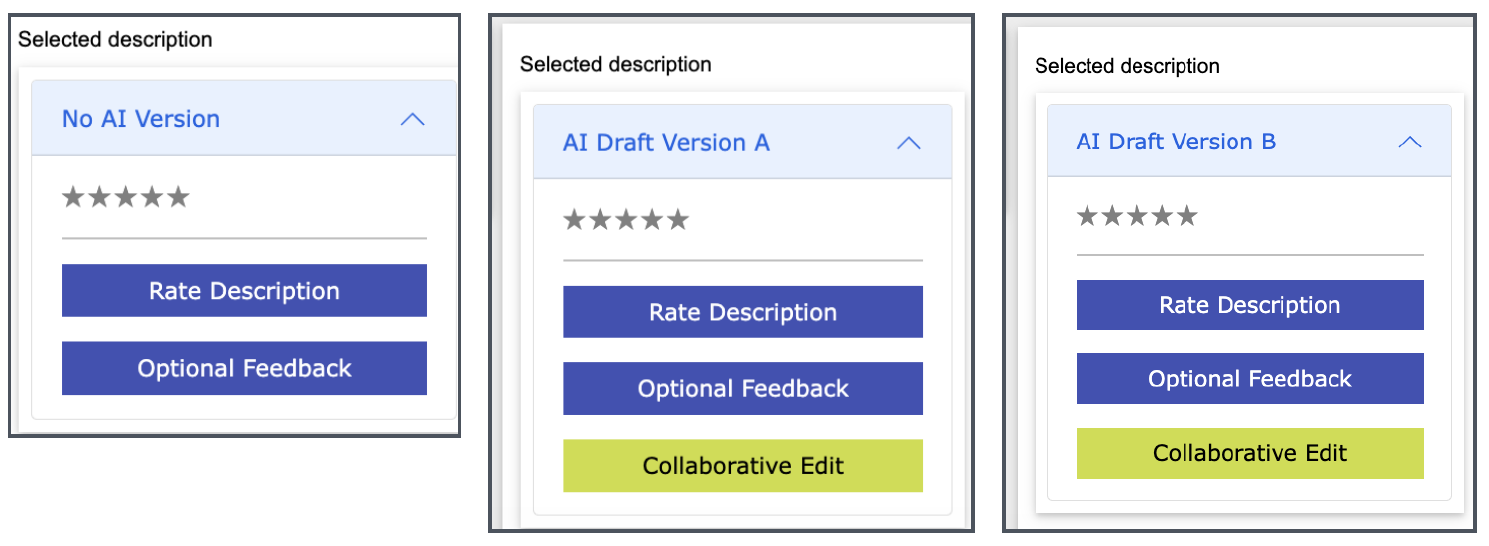}
  \caption{Condition labels presented to participants. The two AI conditions were labeled \textit{AI Draft Version A} and \textit{AI Draft Version B} to avoid signaling differences in generation approach.}
  \label{fig:conditions}
\end{figure}

Table~\ref{tab:study-groups} shows the full condition assignment across groups. Figure~\ref{fig:conditions} shows the anonymized condition labels presented to participants.

\begin{table*}[!htb]
  \centering
  \caption{Condition assignment across four groups. Video order was randomized within each group.}
  \label{tab:study-groups}
  \small
  \begin{tabular}{l ccccc}
    \toprule
    & \textbf{Video 1} & \textbf{Video 2} & \textbf{Video 3} & \textbf{Video 4} & \textbf{Video 5} \\
    & \textit{Crispy Fritters} & \textit{Potato Balls} & \textit{Tangled} & \textit{Neuroscience} & \textit{Origami} \\
    \midrule
    Group 1 & From scratch & GenAD        & Baseline & GenAD    & Baseline \\
    Group 2 & Baseline     & From scratch & GenAD    & Baseline & GenAD    \\
    Group 3 & GenAD        & From scratch & Baseline & GenAD    & Baseline \\
    Group 4 & From scratch     & Baseline & GenAD    & Baseline & GenAD    \\
    \bottomrule
  \end{tabular}
\end{table*}

\FloatBarrier
\section{Model-Assisted Edit Classification Prompts}
\label{appendix:classification-prompts}

We used task-specific prompts for edit classifications requiring
semantic or visual judgments. All model-assisted classifications used GPT-5.6 Sol and constrained responses to predefined output schemas. Programmatic validation checked that returned labels and evidence fields were consistent with the classification rules; invalid responses were retried. The model outputs were subsequently reviewed using the procedure described in Section~\ref{sec:edit-taxonomy}.

The labels shown below are the internal labels used by the analysis scripts. For reporting, \texttt{not\_true} corresponds to
\textit{incorrect}, and \texttt{redundant\_with\_kept\_clip}
corresponds to \textit{redundant with AD}.

\subsection{Deletion Classification}

For each deleted description, the model received the deleted clip and its time span, the video's audio context, nearby descriptions retained in the participant's final track, and timestamped video frames sampled before, during, and after the deleted interval. The prompt imposed a fixed decision order so that \textit{not useful} functioned as a residual category rather than a catch-all.

\noindent
\begin{lstlisting}[style=prompt]
You are classifying why a human editor deleted an
AI-generated audio-description clip from {topic}.

DELETED AD CLIP (span {span}):
"{deleted_text}"

The video's overall soundtrack:
{audio_profile}.

TRANSCRIPT around the deletion
(timestamped; empty means no speech in this window):
{transcript}

RETAINED AD CONTEXT -- the editor's final description
track near the deletion (verbatim, with timestamps):
{kept_context}

You are also given video frames sampled BEFORE the
deleted span, DURING it, and AFTER it, each labeled
with its timestamp. Visual events may not align
exactly with the AD timing; use the before/after
frames, not only the frames during the clip.

Your task: assign ONE primary deletion reason.

CATEGORIES

not_true
The deleted description contains information that is
factually incorrect or unsupported by the video
(e.g., wrong person/object, an action that does not
occur, incorrect attribute/state/expression/spatial
relation, or timing for which the visual evidence
does not support the description). Judge this from
the visual evidence rather than from the text alone.

redundant_with_kept_clip
Another AD clip retained in the edited version already
conveys the same substantive information. Judge
semantic equivalence rather than wording overlap.

redundant_with_audio
The soundtrack (dialogue, narration, sound effects,
or other audible information) already sufficiently
conveys the information.

not_useful
The description is visually supported and is not
redundant with retained AD or audio, but contributes
little to understanding the action, characters,
setting, or narrative. Use this only after ruling out
the other categories.

DECISION ORDER

1. Is the deleted description visually supported?
   If not -> not_true.
2. If supported, does retained AD already convey it?
   If yes -> redundant_with_kept_clip.
3. If not, does the audio already convey it?
   If yes -> redundant_with_audio.
4. Otherwise, if it is low-value in context
   -> not_useful.

Set needs_manual_review=true when the evidence is
ambiguous, the relevant visual event may fall outside
the sampled window, transcript alignment is uncertain,
multiple categories are similarly plausible, or the
judgment requires broader narrative context.

RETURN STRICT JSON:

{
  "primary_label":
    "not_true|redundant_with_kept_clip|
     redundant_with_audio|not_useful",
  "alternative_label": "",
  "confidence": 0.0,
  "visual_supported": true,
  "redundant_with_kept_clip": false,
  "redundant_with_audio": false,
  "useful_if_unique": true,
  "expanded_window": false,
  "visual_evidence": "...",
  "kept_clip_evidence": "...",
  "audio_evidence": "...",
  "reasoning_summary": "...",
  "needs_manual_review": false
}
\end{lstlisting}
\captionof{figure}{Prompt for classifying deletion reasons.}
\label{fig:deletion-class-prompt}

\subsection{Insertion Classification}

Insertion classification focused on semantic novelty rather than
factual correctness. For each insertion, the model received the
participant's inserted description and the complete original AI draft track with timestamps. Thus, information was considered new only when it had not already been conveyed anywhere in the draft provided to the editor.

\noindent
\begin{lstlisting}[style=prompt]
You are classifying what visual information a human
editor added to an AI-generated audio description.

INPUTS:

HUMAN_INSERTION
(placed at {ins_span} in the video):
{hu_text}

AI_CONTEXT
(the complete AI draft description track for this
video, one clip per line as [start-end] text; the
line nearest the insertion is marked with >>):
{ai_context}

TASK:

Identify only the information in HUMAN_INSERTION that
was NOT already conveyed semantically by AI_CONTEXT.
Different wording does not count as new information
if the meaning is equivalent. Information established
earlier in AI_CONTEXT should also be treated as
already known.

Assign zero or more of these labels:

1. scene_setting
   Overall location, environment, time of day, or
   major setting/time transitions.

2. people_identity
   Newly introduced or identified people, names,
   roles, presence/entrance/exit, or distinguishing
   identity information.

3. objects_attributes
   Newly introduced non-human entities or their
   appearance, condition, color, shape, size,
   composition, or other visual properties.

4. actions_expressions
   New actions, events, movements, gestures, body
   language, facial expressions, reactions, or
   visually depicted processes.

5. onscreen_text_data
   New readable text, labels, signs, captions,
   numbers, charts, diagrams, or visually presented
   data.

RULES:

- Multi-label classification is allowed.
- Assign a category only when that information is
  novel relative to AI_CONTEXT.
- Semantically equivalent information expressed in
  different words is already present.
- Attributes can be novel even when their entity is
  already present.
- If the text is too ambiguous to determine its
  information type, set needs_video_review=true
  rather than guessing.
- Do not assess whether the insertion is visually
  correct; this task compares information content.

RETURN STRICT JSON:

{
  "scene_setting": 0,
  "people_identity": 0,
  "objects_attributes": 0,
  "actions_expressions": 0,
  "onscreen_text_data": 0,
  "novel_information": [],
  "already_present_information": [],
  "needs_video_review": false,
  "reasoning": "...",
  "confidence": "high|medium|low"
}
\end{lstlisting}
\captionof{figure}{Prompt for classifying inserted information.}
\label{fig:insertion-class-prompt}

\subsection{Text Revision Classification}

For text revisions, the model directly compared the AI-generated
description with the participant's final version. Classification was based on changes in information content rather than lexical or length differences alone.

\noindent
\begin{lstlisting}[style=prompt]
A novice audio describer revised an AI-drafted audio
description clip for {topic}. Compare the information
conveyed by the two versions and classify the revision.

ORIGINAL (AI draft):
"{ai_text}"

REVISED (final):
"{hu_text}"

Categories (pick exactly ONE):

minor_rewrite
The wording changed without substantially changing
the information conveyed, including paraphrasing,
grammar changes, word choice, or reordering with the
same content.

expanded
The revision retained the original content while
ADDING information.

shortened
Information was REMOVED while preserving the core
description.

major_rewrite
The revision substantially changed the information
conveyed, including content that was replaced,
corrected, or otherwise meaningfully altered. Also
use this when substantial information was both added
and removed.

Judge information content, not wording. A full
rephrasing that conveys the same facts is a
minor_rewrite rather than a major_rewrite. Length
change alone does not determine the category.

RETURN STRICT JSON:

{
  "revision":
    "minor_rewrite|expanded|shortened|major_rewrite",
  "information_added": [],
  "information_removed": [],
  "reasoning": "one sentence",
  "confidence": "high|medium|low"
}

information_added and information_removed should list
the concrete information units added or removed.
For minor_rewrite both must be empty.
For expanded, information_added must be non-empty
and information_removed empty.
For shortened, information_removed must be non-empty
and information_added empty.
For major_rewrite, list the information that changed.
\end{lstlisting}
\captionof{figure}{Prompt for classifying text revisions.}
\label{fig:revision-class-prompt}

\subsection{Visual Resynchronization of Timing Changes}

Playback changes and timing changes explained by dialogue overlap or
description duration were classified deterministically. GPT-5.6 Sol
was used only for residual timing changes for which these explanations
had already been ruled out. The model compared timestamped frames around
the original and edited positions.

\noindent
\begin{lstlisting}[style=prompt]
You are evaluating whether a human editor changed the
timing of an audio-description clip to better
synchronize it with the visual event being described.
The video is {topic}.

Audio-description text:
"{text}"

You are given timestamped frames around the ORIGINAL
position ({old_start}s) and around the EDITED position
({new_start}s).

Audio-overlap and text-duration explanations have
already been evaluated and ruled out. Your task is
ONLY to judge visual synchronization: which timing
position better aligns the description with the
visual event it describes?

Return one of:

ORIGINAL
The original timing aligns more clearly with the
described event.

EDITED
The edited timing aligns more clearly with the
described event.

BOTH
Both positions reasonably align with the same visual
event.

NEITHER
Neither position is well supported by the frames.

UNCERTAIN
The evidence is insufficient or ambiguous.

Judge the visual event, not wording style.

RETURN JSON:

{
  "visual_alignment":
    "ORIGINAL|EDITED|BOTH|NEITHER|UNCERTAIN",
  "confidence": 0.0,
  "visual_evidence": "brief"
}
\end{lstlisting}
\captionof{figure}{Prompt for visual resynchronization of residual timing changes.}
\label{fig:resync-prompt}

\subsection{Deterministic Playback and Timing Rules}

Playback adjustments were classified without a language-model judgment. The analysis used the edited description's actual synthesized-speech duration, dialogue intervals, and temporal overlap with other retained AD clips. An extended-to-inline change was classified as removing an unnecessary pause when the description fit inline without meaningful overlap; an inline-to-extended change was classified as preventing overlap when inline playback would conflict with dialogue or another retained description.

For start-time adjustments greater than 0.5\,s, we applied the timing categories hierarchically. We first tested whether the edited position reduced dialogue overlap. We then tested whether a text revision changed the duration such that the original position could no longer accommodate the revised inline description while the edited position could. Only remaining cases were evaluated for visual-event resynchronization using the prompt above. Changes not supported by these criteria were retained as residual \textit{Other} cases.
\FloatBarrier 
\section{Examples of Edits}
\label{appendix:examples}

\begin{table*}[t]
\centering
\footnotesize
\caption{Representative examples for each edit classification in Figure~\ref{fig:edit-composition}, drawn from the edit logs of both conditions.}
\label{tab:edit-examples}

\begin{tabularx}{\textwidth}{@{}l l X X@{}}
\toprule
Action & Category & Baseline example & GenAD example \\
\midrule

Delete & Redundant w/ AD & ``Surprised blonde in purple.'': deleted; the participant's retained clip ``Blonde woman in purple dress holds a pan, looking curious then startled.'' conveys the same information in more detail. & ``Water is slowly added while stirring until a smooth dough forms'': deleted; the retained clip transcribes the on-screen text ``3/4 cup water. Add gradually.'' \\

Delete & Redundant w/ audio & ``2-MINUTE NEUROSCIENCE'', generated as its own clip: deleted; the narrator speaks the title aloud. & ``A diagram of a neuron is shown. The neuron has branching dendrites extending from the left side, highlighted in yellow. An arrow points from the word 'NEURON' to the dendrites.'': deleted; the narration already conveys this: ``The structures extending from the left side of the neuron that look a little bit like tree branches are called dendrites. Dendrites are the area where neurons receive most of their information.'' \\

Delete & Not useful & ``CAUTION: Hot oil may be released when lid is removed from cookware'': deleted; a safety warning printed on the cookware, unrelated to the recipe steps. & ``Thank you for watching / Please subscribe for more videos'': deleted; a promotional end card. \\

Delete & Incorrect & ``Both corners are folded in, creating a rectangular shape with diagonal creases.'': deleted; the video shows the fold being opened and collapsed inward. & --- \\

\addlinespace

Insert & Actions \& expressions & No draft clip covers Flynn's entry into the tower: ``He slips through the window and shuts it behind him.'' & No draft clip covers the pan spinning after the strike: ``A frying pan was thrown into the air and spun around.'' \\

Insert & People identity & ``Rapunzel tilts Flynn's head with the frying pan to get a look at his face.'': names both characters, which the draft refers to only as ``the woman'' and ``the man.'' & --- \\

Insert & Objects \& attributes & ``A chameleon appears.'': inserted; the draft never introduces the chameleon. & ``Her face showed an expression with sense of achievement and surprise.'': added while Rapunzel talks to herself in the mirror. \\

Insert & Scene setting & ``In a forest, a man with a satchel is watching a vine-clad tower between cliffs\ldots'': establishes the setting of the scene. & --- \\

Insert & On-screen text \& data & ``add 1/2 tbs oil'': transcribes an on-screen measurement omitted from the draft's description of this step. & --- \\

\addlinespace

Change text & Major rewrite & ``A surprised man opened his eyes.'' $\rightarrow$ ``Suddenly, he opens his eyes. In a panic, she raises the frying pan and hits him again.'' & ``A bowl with cabbage and carrot mixture is beside a pan filled with hot oil\ldots'' $\rightarrow$ ``We heat a pan filled with oil. Scoop a spoonful of dough and drop it into the hot oil to get it fried.'' \\

Change text & Shortened & ``The drawing continues with a blue axon extending from the neuron. The hand draws the axon terminals in red\ldots'' $\rightarrow$ ``A hand draws red axon terminals at the end of a blue axon.'' & ``Rapunzel moves slightly closer, her eyes widening. Flynn suddenly opens his eyes.'' $\rightarrow$ ``Rapunzel moves slightly closer, her eyes widening.'' \\

Change text & Expanded & ``A hand places a potato ball into hot oil in a frying pan. The oil bubbles around the potato ball, indicating it's frying.'' $\rightarrow$ ``Put enough olive oil in a pan. Heat the oil until it makes the sound of bubbles. Then put the potato balls one by one into the boiling oil to fry them.'' & ``Oil is added to the mixture, continue to mix till fully incorporated.'' $\rightarrow$ ``Add 1/2 tbs oil to the mixture, continue to mix till fully incorporated.'' \\

Change text & Minor rewrite & ``3/4 cup water / Add gradually'' $\rightarrow$ ``add 3/4 cup water gradually'' & ``In an empty large bowl, 3/4 cup of flour is added.'' $\rightarrow$ ``In an empty large bowl, add 3/4 cup of flour.'' \\

\addlinespace

Playback & Extended $\rightarrow$ inline & ``A person in a green vest climbs in\ldots'': converted to inline; the description fit the gap without pausing playback. & ``Potato balls sizzle as they are placed into bubbling hot oil.'': converted to inline; the description fit the available gap. \\

Playback & Inline $\rightarrow$ extended & ``Crispy vegetable fritters with red sauce on a white plate.'': converted to extended, preventing overlap with the neighboring description. & ``Prepare a pan filled with hot oil on a stovetop.'': converted to extended, preventing overlap with the neighboring description. \\

\addlinespace

Start time & Avoided dialogue & Wardrobe clip overlapping 10.9\,s of dialogue: shifted 4.4\,s earlier, reducing the overlap. & ``Rapunzel startles and swings the frying pan\ldots'' overlapping speech: shifted 3.6\,s, eliminating the overlap. \\

Start time & Resync.\ w/ visual event & ``1 egg'', timed after the egg is already mixed in: shifted 10\,s earlier to when the egg is poured into the flour. & ``Pascal changes from green to red\ldots'' timed ahead of the action: shifted ${\sim}3$\,s later to coincide with it. \\

\bottomrule
\end{tabularx}
\end{table*}
\clearpage

\section{Interactive Onboarding Tutorial}
\label{appendix:tutorial}

\begin{figure}[!htb]
  \centering
  \includegraphics[width=\textwidth]{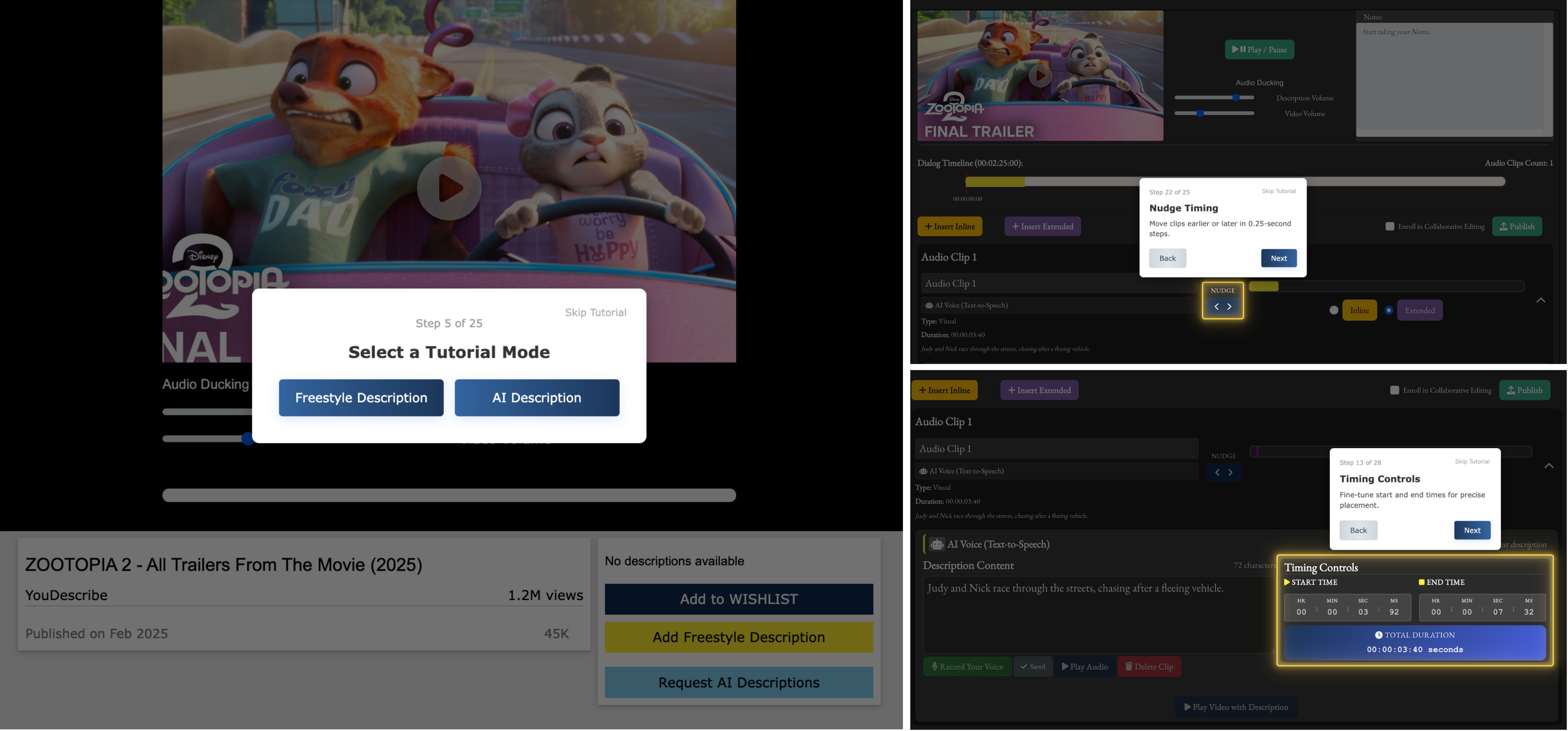}
  \caption{Interactive onboarding tutorial. A spotlight overlay guides users step by step through the interface, requiring specific interactions before advancing. At an intermediate stage, the tutorial branches into manual authoring or AI draft refinement paths.}
  \label{fig:tutorial}
\end{figure}

Based on participant feedback that the full set of editing features felt overwhelming on first use, we implemented an interactive tutorial mode (Figure~\ref{fig:tutorial}) to onboard new users before they begin describing.

\end{document}